\documentclass{aa}  

\usepackage{graphicx}
\usepackage{txfonts}
\usepackage{ulem}
\usepackage{hyperref}

\providecommand{\spins}{{\sc SPInS}}
\providecommand{\flame}{{\sc Flame}}
\providecommand{\apsis}{{\sc Apsis}}
\providecommand{\basti}{{\sc BaSTI}}
\providecommand{\algoone}{{\sc Algo1}}
\providecommand{\algotwo}{{\sc Algo2}}
\providecommand{\gspphot}{{\sc GSP-Phot}}
\providecommand{\gspspec}{{\sc GSP-Spec}}
\providecommand{\teff}{\ensuremath{T_{\rm eff}}}
\providecommand{\msol}{M\ensuremath{_{\odot}}}
\providecommand{\lsol}{L\ensuremath{_{\odot}}}
\providecommand{\rsol}{R\ensuremath{_{\odot}}}
\providecommand{\rad}{\ensuremath{R}}
\providecommand{\lum}{\ensuremath{L}}
\providecommand{\mass}{\ensuremath{M}}
\providecommand{\age}{\ensuremath{\tau}}
\providecommand{\evol}{\ensuremath{\epsilon}}
\providecommand{\logg}{\ensuremath{\log g}}
\providecommand{\mh}{\ensuremath{\rm{[M/H]}}}
\providecommand{\feh}{\ensuremath{\rm{[Fe/H]}}}
\providecommand{\alphafe}{\ensuremath{\rm{[\alpha/Fe]}}}
\providecommand{\ag}{\ensuremath{A_G}}
\providecommand{\av}{\ensuremath{A_V}}
\providecommand{\mbolsol}{\ensuremath{M_{\rm bol,\odot}}}

\providecommand{\rvgrav}{\ensuremath{V_{\rm GR}}}
\usepackage{tikz}

\usepackage{pifont}
\usepackage{siunitx}
\usepackage{acronym}
\acrodef{gdr}[GDR]{Gaia Data Release}
\acrodef{pic}[PIC]{Plato Input Catalogue}
\acrodef{snr}[SNR]{Signal-to-Noise Ratio}
\acrodef{mad}[MAD]{Median-absolute-deviation}

\begin{document}

   \title{Stellar masses and ages in Gaia Data Release 4 from the Final Luminosity Age Mass Estimator algorithm}

\titlerunning{Masses and ages in GDR4}
\authorrunning{Creevey et al.}
   \author{
   O.~L.~Creevey
          \inst{1}\fnmsep\thanks{orlagh.creevey@oca.eu}
          \and
          L.~Casamiquela\inst{2}
\and
Y.~Lebreton\inst{2,3}
\and
C. Ordenovic\inst{1}
\and
F. Th\'evenin\inst{1}
\and
M. Fouesneau\inst{4} \and
R.~Andrae\inst{4} \and 
C.~A.~L.~Bailer-Jones\inst{4}  \and 
F.~Pailler\inst{5}
\and
B. Pichon\inst{1} \and 
C. Babusiaux\inst{6} \and 
A.~Barbier\inst{5} 
\and
N. Baudeau\inst{5,7} 
\and 
N. Brouillet\inst{8}  
\and
S. Cassisi\inst{9,10} 
\and
B. Edvardsson\inst{11}  
\and
O. Kochukhov\inst{11}  
\and 
G. Kordopatis\inst{1}
\and
A.~C. Lanzafame\inst{12,13} 
\and
A. Korn\inst{11} 
\and 
S. Meszaros\inst{14,15,16}
\and
C. Navarrete\inst{1,5} 
\and
C.~Robin\inst{5}
\and
R. Sordo\inst{17} 
\and
C. Soubiran\inst{8} 
\and
D.~R. Reese\inst{2} 
          }

   \institute{Université Côte d'Azur, Observatoire de la Côte d'Azur, CNRS, Laboratoire Lagrange, Bd de l'Observatoire, CS 34229, 06304 Nice cedex 4, France\\
              \email{orlagh.creevey@oca.eu}
         \and
LIRA, Observatoire de Paris, Université PSL, Sorbonne Université, Université Paris Cité, CY Cergy Paris Université, CNRS, 92195, Meudon, France
\and
        Université Rennes, CNRS, IPR (Institut de Physique de Rennes) - UMR 6251, F-35000 Rennes, France
    \and
Max Planck Institute for Astronomy, Königstuhl 17, 69117 Heidelberg, Germany
             \and
Centre national d’études spatiales (CNES), 2, Place Maurice Quentin, 75039, Paris, France
\and
Universit\'e Grenoble Alpes, CNRS, IPAG, 38000 Grenoble, France
\and
Telespazio France for CNES Centre Spatial de Toulouse, 18 avenue Edouard Belin, 31401 Toulouse Cedex 9, France           
\and
    Laboratoire d'Astrophysique de Bordeaux, Univ. Bordeaux, CNRS, B18N, all\'ee Geoffroy Saint-Hilaire, 33615 Pessac, France
             \and             
INAF-Osservatorio Astronomico d'Abruzzo, via M. Maggini, sn. 64100, Teramo, Italy
\and
INFN, Sezione di Pisa, Largo Pontecorvo 3, 56127 Pisa, Italy
\and
Department of Physics and Astronomy, Uppsala University, Box 516, SE-751 20 Uppsala, Sweden
\and
Dipartimento di Fisica e Astronomia "Ettore Majorana", Università di Catania, Via S. Sofia 64, 95123, Catania, Italy
\and
INAF - Osservatorio Astrofisico di Catania, Via S. Sofia 78, 95123, Catania, Italy
\and
ELTE E\"otv\"os Lor\'and University, Gothard Astrophysical Observatory, 9700 Szombathely, Szent Imre H. st. 112, Hungary
\and
MTA-ELTE Lend{\"u}let "Momentum" Milky Way Research Group, Hungary
\and
HUN-REN CSFK, Konkoly Observatory, Konkoly Thege Mikl\'os \'ut 15-17, Budapest, 1121, Hungary
\and
INAF-Osservatorio Astronomico di Padova, Vicolo Osservatorio 5, 35122 Padova, Italy
}

   \date{Received May 22, 2026}

  \abstract
  {The masses and ages of stars are key quantities for understanding exoplanetary, stellar, and galactic evolution. In the context of Gaia, these parameters provide insights into the stellar populations, helping to trace the formation and history of the Galaxy.} 
   {As part of the Gaia Data Processing and Analysis Consortium (DPAC), the Final Luminosity Age Mass Estimator (FLAME)  pipeline processes Gaia data to derive stellar parameters comprising luminosities, radii, masses and ages. This paper discusses the methods and data used in \flame\ for Gaia Data releases and the expected performances of FLAME for the 4th Gaia Data Release.}
   {FLAME comprises two main components: the first one, which is analytical, is used to estimate luminosity, radius, and radial velocity correction due to gravitational redshift by exploiting the atmospheric, astrometric, and photometric parameters produced within Gaia.  The second is a model inference based on two main approaches: a classical minimization approach, and a Bayesian framework.  It aims to derive mass, age, and evolutionary stage.  The two step implementation offers flexibility in handling photometric properties that are prone to systematic errors.
   }
   {Tests with simulated data, the Sun, and well characterised samples of stars including some clusters show that the methods in \flame\ perform as expected, producing results in statistical agreement with the literature.  Mass and age for giant stars are extremely sensitive to the input atmospheric parameters and good agreement with external data for these stars is only reached when we use the same input data.  This emphasizes the difficulty in validating masses and ages in any catalogue.  
   We provide new stellar fundamental parameters for some high velocity stars, stars with very low mass companions, and a selection of stars in the Plato Field of View.
      We also discuss the expected uncertainties and show that typical mass uncertainties are in relatively good agreement with analytical predictions from a mass-luminosity relation.   Typical relative uncertainties on age for solar-like stars vary between 20\% -- 40\%, and for masses between 1.2 and 1.3 \msol\ they are around 20\%.  For giants, the relative uncertainties in age are typically 10\% for lower masses, increasing to 15--20\% for higher masses.   
   }
   {We conclude that \flame\ produces valid results in Gaia Data Releases.  In Gaia Data Release 4 approximately 500 million sources have results from the pipeline, thus providing large samples of stars with high precision and accuracy in their stellar parameters.  As the results depend on the quality of the input data, samples with a much decreased quality are also expected.  Users should consult the Gaia online documentation and flags for guidelines on the exploitation of the catalogue.
}


   \maketitle
\nolinenumbers

%

\section{Introduction}\label{sec:intro}
The age of a star is a fundamental parameter for understanding the formation and evolution of astrophysical systems. Reconstructing the history of our Milky Way, which has been shaped by the accretion of smaller galaxies \citep{searle78,helmi18}, requires determining the ages of stars belonging to its various populations \citep{helmi20,gallart19}, as revealed by Gaia \citep{gaiadr1}. Furthermore, for individual stars, estimating the age of a star that hosts an exoplanet is essential for constraining the evolutionary history of exoplanetary systems, a goal that will be further advanced by the upcoming PLATO mission \citep{plato2025}. For the scientific return of both space missions, stellar mass is a fundamental parameter, as it governs the evolution of a star over time. 

The Gaia mission \citep{gaiamission}, launched in 2013 by the European Space Agency (ESA), aimed to produce a comprehensive six-dimensional map of the Milky Way by measuring the precise positions and velocities of approximately 1\% of the Galaxy’s stars \citep{gaiadr1,gaiadr2,gaiadr3}. In addition to mapping the Galaxy \citep{rowell21,torra21}, Gaia provides detailed physical characterisation of its observed sources using high-precision astrometry \citep{lindegren21}, photometry \citep{riello21}, and low- \citep{francesca23} and medium-resolution spectroscopy  \citep{sartoretti2018,sartoretti2023,katz2023}. The determination of stellar masses and ages is part of this characterisation effort, carried out by the Gaia Data Processing and Analysis Consortium (DPAC), specifically through Coordination Unit 8 (CU8 -- Astrophysical Parameters) pipeline named Apsis \citep{apsis,apsisdr2,apsisdr3paper1,apsisdr3paper2,apsisdr3paper3}.

{ Both the mass and age of stars are derived by the module \flame\ within the \apsis\ workflow. The \flame\ module is designed to compute the radii, luminosities, radial velocity correction due to gravitational redshift, masses, ages, and evolutionary stages of all processed sources. In Gaia Data Release 3 (GDR3, June 2022), the \flame\ results are provided in the {\tt astrophysical\_parameters} and {\tt astrophysical\_parameters\_supp} tables.
The goal of this article is to describe the \flame\ module and its resulting products, as well as to evaluate its performance in preparation for Gaia Data Release 4 (GDR4). 

The article is organised as follows. The architecture of the module and the methodology are described in Sect.~\ref{sec:method}, together with the models implemented in the analysis. The Gaia data used by \flame\ are presented in Sect.~\ref{sec:data}. The validation of the methodology based on simulated data and the Sun is discussed in Sect.~\ref{sec:validation}, while  Sect.~\ref{sec:compare} extends this validation by comparing \flame\ results with those from other catalogues.
In Sect.~\ref{sec:apply}, we apply \flame\ to infer the stellar properties—in particular masses and ages—for several samples of stars of interest.
In Sect.~\ref{sec:specs}, we finalise the work by discussing the expected uncertainties in \flame\ parameters in GDR4, before concluding in 
Sect.~\ref{sec:conclude}.



\section{Method and Models}\label{sec:method}
The \flame\ code\footnote{\url{https://indico.ict.inaf.it/event/2345/}} \citep{creeveytnote,gdr3onlinedoc-cu8,apsisdr3paper1} derives stellar evolutionary parameters and mass by exploiting atmospheric parameters, astrometry, photometry and stellar  models. It comprises two components: \algoone\, which estimates stellar luminosity \lum, radius \rad, and radial velocity correction for gravitational redshift \rvgrav, and \algotwo\, which infers model parameters using the derived \lum, and complementary atmospheric parameters.
For both components it exploits atmospheric parameters produced within \apsis\ (see Section~\ref{sec:data}).
The choice to separate the \flame\ algorithm into an observational module and a model-based inference component is twofold.
Firstly, it enables the encapsulation of all complexities of the photometric bands into a single parameter — the bolometric correction.
Secondly, it allows for a direct comparison of the derived physical properties of the stars with the true theoretical ones predicted by the stellar structure equations, namely the luminosity, \teff, and \logg.
Additionally, this separation facilitates both technical and scientific validation of the method based on observational data obtained from other observatories, enabling straightforward and direct comparisons between the resulting parameters.

\subsection{\algoone}

{The first algorithm in \flame, \algoone, computes the luminosity, \lum, radius, \rad, and the radial velocity correction due to gravitational redshift, \rvgrav. 
The luminosity is calculated using the equation}
\begin{equation}
    M_{\rm bol} - M_{\rm bol, \odot} = -2.5 \log_{10} \left ( \frac{L}{L_\odot} \right ),
    \label{eqn:def-luminosity}
\end{equation}
where \begin{equation}
M_{\rm bol} =  G  - 5 \log_{10}d + 5 - A_G + BC.
\label{eqn:def-magbol}
   \end{equation} 
Here, $G$ is the Gaia $G$-band magnitude, \ag\ is the extinction in the $G$-band, $d$ is the distance in parsecs, and BC is the bolometric correction in $G$ band, which depends on the atmospheric parameters \teff, \logg, \feh, and \alphafe\footnote{\feh\ is the logarithm of iron-to-hydrogen abundance ratio relative to the Sun, and \alphafe\ is the $\alpha$-element-to-iron ratio}. We adopt \mbolsol\ = 4.74 mag \citep[IAU 2015 resolution;][]{mamajek15}.
Within \flame, $G$ is calculated from the observed flux $f_G$, its uncertainty, and the photometric zeropoint of the passband, allowing for accurate propagation of uncertainties.

Once \lum\ is determined, the radius \rad\ is calculated from the Stefan-Boltzmann relation with the same \teff\ used for the determination of the BC.  
\begin{equation}
{R/R_\odot = \frac{(L_\star{/L_\odot})^{1/2} }{(\teff / \teff{_{\odot}} )^2}}
\label{eqn:stefan-boltzmann}
\end{equation}
The radial velocity correction to the gravitational redshift is calculated from the input \logg\ and the derived $R$, 
\begin{equation}
 \rvgrav =   \frac{GM}{Rc} =   \frac{GM}{R^2} \frac{R}{c} = g \frac{R}{c},
\label{eqn:gravshift}
\end{equation}
where $c$ is the speed of light in vacuum.

Parameter values and their uncertainties are estimated using two methods: the first a bootstrap method, where the input observations (\teff, \logg, \feh, \alphafe, $f_G$, $d$, \ag) are perturbed $N$ times to generate distributions of (\lum, \rad, \rvgrav). The second method uses the Markov Chain Monte Carlo samples from the preceding module (\gspphot).
The resulting median of each distribution defines the parameter value, while the 0.16 and 0.84 quantiles provide the lower and upper bounds of the 68\% confidence interval.

\subsection*{Bolometric correction}\label{ssec:bc}
The BC, in the $G$-band, is evaluated on a grid of stellar atmosphere models using the relevant passbands and zeropoints.  For \ac{gdr} 3 the passbands\footnote{\url{https://www.cosmos.esa.int/web/gaia/dr3-passbands}} and the bolometric corrections\footnote{\url{https://www.cosmos.esa.int/web/gaia/dr3-bolometric-correction-tool}} can be retrieved from the ESA cosmos website.  The grids of models and methods used for calculating the BC within Gaia-DPAC \apsis\ modules are described in detail in \citeauthor{kornmodels}. Each model requires the input parameters {\teff, \logg, \feh, \alphafe} and returns BC values in the $G$, $G_{BP}$, and $G_{RP}$ bands. For \teff\ in the range 2500–8000 K, we use the MARCS models \citep{marcsmodels}, while for \teff\ above 8000 K, Line-by-Line opacity stellar models (typically referred to as LL or A models) are adopted \citep{llmodels}. The transition at 8000 K is handled by matching each set of models defined by {\alphafe, \feh, \logg} in \teff, ensuring a smooth connection between the two grids.

{The models include a zeropoint offset that must be corrected, effectively defining BC$_{G,\odot}$.   In \ac{gdr}3, this was achieved by defining $M_{G,\odot} = 4.66$ mag by using several independent methods (models, external datasets, and solar analogues), which then yielded BC$_{G,\odot} = 0.062$ mag \citep{apsisdr3paper1}, with an estimated uncertainty of ±0.015 mag.  In \ac{gdr}4, these values are revised based on the DR4 photometric passbands and will be published in the online documentation and \apsis\ papers.}

\begin{table*}[]
    \caption{Parameters published in the Gaia archive from \flame.}
    \centering
    \label{tab:outputflame}
    \begin{tabular}{lllll}
    \hline\hline
    Parameter & description& XP & RVS\\
    \hline
    
\lum  & luminosity & {\tt lum\_flame} &  {\tt lum\_flame\_spec} & \\
\rad & radius & {\tt radius\_flame}  & {\tt radius\_flame\_spec}& \\
\rvgrav & radial velocity correction due & {\tt gravredshift\_flame}  & {\tt gravredshift\_flame\_spec} & \\
& to the gravitational redshift\\
BC  & bolometric correction & {\tt bc\_flame} &  {\tt bc\_flame\_spec} & \\
\\
\mass & mass&{\tt mass\_flame}  & {\tt mass\_flame\_spec} & \\
\age & age & {\tt age\_flame} &  {\tt age\_flame\_spec}& \\
\evol & evolution stage & {\tt evolstage\_flame}  & {\tt evolstage\_flame\_spec}& \\
\mh & model metallicity&{\tt mh\_flame\_model} &  {\tt mh\_flame\_model\_spec}& \\
\logg & model \logg& {\tt logg\_flame\_model} &  {\tt logg\_flame\_model\_spec}& \\
\teff & model \teff & {\tt teff\_flame\_model} &  {\tt teff\_flame\_model\_spec}& \\
\lum  & model \lum & {\tt lum\_flame\_model} &  {\tt lum\_flame\_model\_spec}& \\
    \hline\hline
    \end{tabular}
    \tablefoot{XP and RVS refer here to the output parameters based on the input atmospheric parameters using XP and RVS spectra, respectively.
    In GDR3 the \flame\ XP-based results are available in the {\tt astrophysical\_parameters} table while the \flame\ RVS-based results are found in the {\tt astrophysical\_parameters\_supp} table.
    In GDR4 the table structure of the astrophysical parameters has been updated and the \flame\ results will be available in the {\tt ap\_xp} and {\tt ap\_rvs}, respectively.}
\end{table*}

\subsection{\algotwo}
The second component of \flame\ performs model inference by comparing the observed stellar properties with their model-based counterparts. This allows the inference of the stellar mass (\mass), age (\age), and evolutionary stage (\evol) {where \evol\ is a number typically between 100 and 1600, with values $<420$ indicating main sequence, and $>490$ from the base of the red giant, see \citet{Hidalgo2018} for details, or the data model \href{https://gea.esac.esa.int/archive/documentation/GDR3/Gaia_archive/chap_datamodel/sec_dm_astrophysical_parameter_tables/ssec_dm_astrophysical_parameters.html#astrophysical_parameters-evolstage_flame}{here}}.
Unlike in \ac{gdr}3, \ac{gdr}4  additionally provides the corresponding model parameters for \teff, \lum, \logg, and \mh, where [M/H] is the ratio of total metal to hydrogen abundance observed in the atmosphere.

\subsubsection{Parameter inference}
 The model-based inference, \algotwo, comprises two methods.  The results published in \ac{gdr}4 are obtained by exploiting the method modelled on the Stellar Parameters INferred Systematically code \citep[\spins, ][]{spins}, a publicly available Python pipeline\footnote{\url{https://gitlab.obspm.fr/dreese/spins}}. The code employs a Bayesian framework to provide the posterior probability distribution function
 (PDF) of the inferred stellar parameters from a set of observational constraints, a grid of stellar models, and a set of priors. The PDF is sampled using a Markov Chain Monte Carlo (MCMC) solver built on the emcee Python package \citep{foremanmackey2013},
 coupled with an interpolation scheme for the stellar models. \flame\ uses the version of the \spins\ code described in detail in \cite{casamiquela2024},  which includes several improvements, such as the on-the-fly computation of the autocorrelation time and the automatic convergence evaluation.
 \spins\ allows the use of priors on the grid parameters, such as the initial mass function (IMF), the metallicity distribution function, or the star formation rate (SFR), see \cite{spins} for details. 
 {We use a flat prior on the age with range [0,13.8] Ga, and the IMF of \citet{salpeter55} and further details on the priors for DR4 are discussed in the online documentation.}
 SPInS has been adapted in Java to be implemented in the CU8/DPAC pipeline at CNES}.

The observational input to the model-parameter inference comprises the set (\lum, \mh, \teff, \logg) and their uncertainties.   The uncertainties are taken as half of the difference between the upper (84\%) and lower (16\%) percentiles for each input parameter.
The results of the model-parameter inference are MCMC samples, and we use the median, 16$^{\rm th}$ and 84$^{\rm th}$ percentiles, after elimination of the burn in, to define the model property and confidence interval.   The model properties that are published are \mass, \age, \evol, and the associated model parameters (\lum, \teff, \mh, \logg), along with their confidence intervals.  We additionally publish the median of the log-posterior value, as an indicator of the quality of the fit, which can be used in conjunction with the model \teff, \lum, \logg, and \mh, to judge the fit.

A second inference method was also implemented in \flame\ and is based on a Levenberg-Marquardt approach \citep{creeveytnote}.    In \ac{gdr}{3} the model-based parameters were derived using this implementation, which provides similar results to \spins\ when the dataset are of high quality.  Additionally in \ac{gdr}{3} a narrow prior based on solar metallicity was implemented due to some known issues on the input metallicity, and empirical corrections to the masses for a non-solar metallicity star are available in \cite{creeveytnote}\footnote{Gaia Public Documents are found \url{https://www.cosmos.esa.int/web/gaia/public-dpac-documents}. }.

\subsubsection{Stellar models}

The stellar models employed in this work are taken from the BaSTI\footnote{\url{http://basti-iac.oa-abruzzo.inaf.it}} grid of stellar models \citep{Hidalgo2018} These models are computed assuming a solar-scaled heavy-element distribution and include key input physics such as atomic diffusion of helium and metals, convective-core overshooting, and mass loss.
The adopted solar mixture follows \citet{Caffau2011}, complemented by the element abundances from \citet{Lodders2010}. Evolutionary tracks are available for initial stellar masses in the range {$M_0 \in [0.1\,M_{\odot},15\,M_{\odot}]$ } and metallicities [M/H] $\in [-3.197,+0.45]$.
We additionally include models at [M/H] = +0.45 without diffusion, since this metallicity is  absent in the BaSTI diffusive grid due to the lack of suitable radiative opacity tables. However, at a so high  metallicity, the extension of the outer convection zone is so large, that the effect of atomic diffusion is hugely reduced - if any - with respect to more metal-poor stars. Moreover, only a small fraction of stars in our sample is affected, making this approximation quite reasonable.

The initial helium abundance follows a helium-to-metal enrichment ratio of $\Delta Y / \Delta Z = 1.31$ \citep[see][for details]{Hidalgo2018}. For each stellar model defined by a given age, initial mass, and chemical composition, the grid provides the current stellar mass, evolutionary stage, luminosity, and effective temperature. From these quantities, the radius, surface gravity, and mean density can be derived. The likelihood function used for model inference is based on the observables $(L,\teff,\logg,\mh)$.

The BaSTI grid spans all evolutionary phases from the pre-main sequence up to either the first thermal pulses on the asymptotic giant branch (AGB), or carbon ignition, or the age of the Universe, depending on stellar mass.
For use in \flame, we constructed a dedicated sub-grid by excluding high-mass models ({$M > 10\,M_{\odot}$}) and retaining only the evolutionary stages from the zero-age main sequence (ZAMS) to the tip of the red giant branch.
This sub-grid comprises a total of \num{16622293} stellar models.

For most stars, the global metallicity derived in the Gaia analysis is from MARCS model atmospheres \citep{marcsmodels}. The MARCS grid adopts the solar abundance mixture of \citet{2007SSRv..130..105G}, whereas the BaSTI models are based on the solar composition of \citet{Caffau2011}, complemented by \citet{Lodders2010}. These two mixtures differ in their overall metallicity scale, with the \citet{Caffau2011} mixture being higher by 0.0976 dex relative to that of \citet{2007SSRv..130..105G}. To ensure consistency between the observational inputs and the BaSTI model grid, we therefore apply a correction of this value to the estimated \mh\ values.

The metallicity [M/H] is the decimal logarithm of the ratio of the total number abundances  of metals to hydrogen with respect to the Sun. If the assumption of a universal solar-scaled mixture is valid, then $\mathrm{[M/H]} \approx \mathrm{[Fe/H]}$. However, when a star’s mixture is enriched in $\alpha$-elements (O, Ne, Mg, Si, S, Ar, Ca, Ti) with respect to the Sun, that is $\mathrm{[\alpha/Fe]} \ne 0$, this approximation is no longer valid. The correct way to infer the parameters of such a star would then be to interpolate within several sets of grids of stellar evolutionary tracks based on different $\alpha$-element enhancements. 
{BaSTI grids are also available for an enrichment of $[\alpha/\mathrm{Fe}]=+0.4$ , however using it in \flame\  would require $[\alpha/\mathrm{Fe}]$ as an additional parameter in the \spins\ optimisation which would considerably increase computation time. On the other hand, }\citet{Salaris+1993} showed that $\alpha$-enhanced isochrones are well-mimicked by standard solar-scaled ones if metallicity is rescaled according to $\alpha$-element enrichment. 
For the \citet{Caffau2011} solar mixture used in the BaSTI grids, we inferred a correction of the form 
\begin{equation}
 {\mathrm{[M/H]}_\alpha \approx \mathrm{[Fe/H]} + 0.76 [\alpha/\mathrm{Fe}]},
    \label{eqn:alphacorrection}
\end{equation}
or in quadratic form
\begin{equation}
{\mathrm{[M/H]}_\alpha \approx \mathrm{[Fe/H]} + 0.208 [\alpha/\mathrm{Fe}]^2+ 0.655[\alpha/\mathrm{Fe}].}    
\label{eqn:alphacorrection2}
\end{equation}
For the stars with non-zero \alphafe\ we corrected the metallicity using Eq.~\ref{eqn:alphacorrection2}.


\section{Gaia data}\label{sec:data}
As part of the Gaia-DPAC system, \flame\ can only use Gaia data and Gaia-derived parameters as input.  
The module produces two sets of results, one based on the  atmospheric parameters derived from the \apsis\ module \gspphot\ \citep{gspphot}, which analyses the Gaia BP and RP spectra (together known as XP) and are available for the vast majority of Gaia sources, and one based on the atmospheric parameters from the \apsis\ module \gspspec\ \citep{gspspec}, which analyses the high resolution RVS spectra for a set of several million sources with a high enough \ac{snr} to derive reliable parameters.    In \ac{gdr}3 results for approximately 300 million sources with $G<18.25$ mag were produced by processing the results from \gspphot, and these are found in the {\tt astrophysical\_parameters} table, while results for approximately 6 million sources with $G<13$ were produced by processing the input atmospheric parameters from \gspspec, and these are found in the 
 {\tt astrophysical\_parameters\_supp} table.   
 All results derived using \flame\ contain {\tt flame} in their parameter name, 
 e.g. {\tt mass\_flame} in the former table or {\tt radius\_flame\_spec\_lower} in the latter where here lower means the 16 percentile value, see Table~\ref{tab:outputflame}.
For DR4 we expect results on approximately 500 million sources based on the \gspphot\ atmospheric parameters and 34 million sources based on the \gspspec\ ones.

The processing by \flame\ is done on a source-by-source basis.  This implies that it has no knowledge if the object is a member of a binary system or a stellar cluster.  Indeed it can also be a non-stellar source, and while we could publish a meaningless radius and luminosity for these sources, it is not expected that the model parameters converge and may be therefore absent. As additional constraints cannot be placed on the processing of the sources, there are evidently parameter spaces that may not perform well, e.g. the ages of very low mass stars or binaries with non-negligible secondary mass components.  We choose to not remove these parameters from the archive as they may in any case provide some important information to the user in \ac{gdr}4. 

\subsection{Parameters from \gspphot}
\gspphot\ processes the low resolution XP spectra \citep{francesca23} to derive the atmospheric parameters (\teff, \logg, and \mh)  and interstellar extinction in the $G$-band, \ag.  The low-resolution spectra span approximately 300--1000 nm and have a typical resolving power of between 30-70 \citep{francesca23}.  For not very high SNR the \teff\ and \ag\ can become degenerate, as they both produce similar changes in the spectral shape of the source\footnote{As an example of the degeneracy between \teff\ and \ag, see the illustration in Fig.~1 on the Gaia webpage stories webpage \url{https://www.cosmos.esa.int/web/gaia/dr3-what-is-in-between-the-stars}}.  \gspphot\ also exploits the parallax and estimates a distance using priors, and this along with the amplitude of the XP spectra provides an independent estimate of the stellar radius.  We note that \gspphot\ employs stellar isochrones to help constrain the solution for the atmospheric parameters, and therefore a mass and age is associated with the optimal parameters.    Full details of \gspphot\ processing can be found in \cite{gspphot}.

The APs from \gspphot\ are produced by several stellar libraries (four libraries in Gaia DR3, three in Gaia DR4): MARCS, PHOENIX, A (LL), and OB \citep{kornmodels}, and while all of the results are available in the Gaia releases in the {\tt astrophysical\_parameters\_supp} table, \flame\ only processes the {\it best} model for each source.  These are found in the {\tt astrophysical\_parameters} table and the best models are typically MARCS for FGK stars, and A for intermediate temperature stars (OB stars are outside of the \flame\ validity range and are therefore not processed).   
In \ac{gdr}4 the \gspphot\ {\it best} and \flame\ results based on XP spectra will be in a table called {\tt ap\_xp}.

\gspphot\ is based on a MCMC approach.  In \algoone\ \flame\ processes the samples provided in the MCMC by \gspphot.
Table~\ref{tab:inputgaiadata} summarises the input data to \flame.

\subsection{Parameters from \gspspec}
\gspspec\ processes the high resolution  RVS spectra to produce atmospheric parameters and chemical abundances. 
The RVS spectra span {wavelengths in the range} 840 -- 870 nm and have a typical resolving power of $R\sim$ \num{11000} \citep{sartoretti2018}. 
As the RVS spectra are not sensitive to interstellar extinction and therefore do not provide this information, we use \ag\ provided by \gspphot\ in Eq.~\ref{eqn:def-magbol}, while we rely uniquely on the parallax as the distance estimator. 

Two sets of results from \gspspec\ are available in the Gaia archive; one based on the Matisse-Gauguin algorithm \citep{bijaoui10,gspspec0,gspspec} and a second one based on a neural-network (ANN).
\flame\ processes the results produced by the Matisse-Gauguin algorithm which are found in the {\tt astrophysical\_parameters} table in \ac{gdr}3, but the \flame\ results are made available in the {\tt astrophysical\_parameters\_supp} table.  
For GDR4 the \gspspec\ and RVS-based \flame\ results are available in a table called {\tt ap\_rvs}.

\subsection{Parameters from astrometry and photometry}
In addition to the atmospheric parameters, \ag, and $d$ provided by the \apsis\ modules, \flame\ also uses the parallax, $\varpi$, from the astrometric solution and the $G$-band flux from the photometric-processing system.  
Within \apsis, the parallax bias \citep{lindegren21} is also applied.

\begin{table}[]
    \caption{Input Gaia data to \flame.} 
    \label{tab:inputgaiadata}
    \centering
    \begin{tabular}{lll}
    \hline\hline
    & XP & RVS\\
    \hline
        & $d$ & $\varpi$ \\
         & $G$ & $G$ \\
        & \teff$_{\rm XP}$ & \teff$_{\rm RVS}$ \\
       &  \logg$_{\rm XP}$ & \logg$_{\rm RVS}$ \\
       &  \mh$_{\rm XP}$ & \mh$_{\rm RVS}$ \\
       & --  & \alphafe$_{\rm RVS}$ \\
       &  \ag$_{\rm XP}$ & \ag$_{\rm XP}$ \\
    \hline\hline
    \end{tabular}
    \tablefoot{XP refers to the parameters provided by \gspphot\ which processes the Gaia XP spectra, while RVS refers to those provided by \gspspec\ which processes the Gaia RVS spectra.}
\end{table}


\section{Validation of the methodology\label{sec:validation}}
In this section, we focus on validating the methodology used in \flame. While \algoone\ is uniquely concerned with the application of analytical equations, our attention here is on the performance of \algotwo\, which handles model-based parameter inference.
To achieve this, we assume an input observational dataset consisting of (\lum, \teff, \logg, \mh), with various expected errors. We then apply \flame\ \algotwo\ to recover the stellar model parameters (\mass, \age, \evol), along with the corresponding model properties (\lum, \teff, \logg, \mh).

\subsection{Simulated data from the stellar grid \label{sec:valid-grid}}
\begin{figure}
    \centering
    \includegraphics[width=0.95\linewidth]{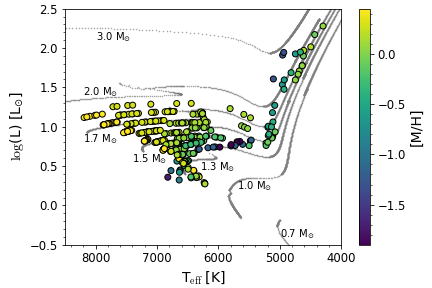}
    \caption{HR diagram showing the sample of simulated data used for validating the methodology.   The background grey tracks are solar-metallicity evolution tracks for different initial mass values.}
    \label{fig:simulateddataHR}
\end{figure}

{We begin with tests using simulated data from the model grid that is used for the inference.  This allows us to perform a direct comparison of the input-output.
We selected 372 stars across the HR diagram, ensuring coverage along the main sequence (MS), sub-giant phase (sG), and red giant branch (RGB), at all metallicities.   A first set was selected using points directly on the model grid, and then a second set was selected using datapoints between model grid points.  
Fig.~\ref{fig:simulateddataHR} shows the distribution of the simulated data set (similar for both cases) across the HR diagram, colour-coded by \mh, with a few solar-metallicity evolution tracks highlighted in the background.  
We varied the input uncertainties from between 1 -- 5 \% errors on \teff\ and \lum, but here we focus on one set of results with 2\% errors in these properties, while fixing the errors in \logg\ and \mh\ 
to 0.1 dex, i.e. we perturbed the model observables by an amount corresponding to their uncertainties and performed the model fitting with these.  The 2\% 
is a typical uncertainty on the input parameters for bright Gaia sources, e.g. $G<14$.}

\begin{figure*}
    \centering
    \includegraphics[width=0.43\linewidth]{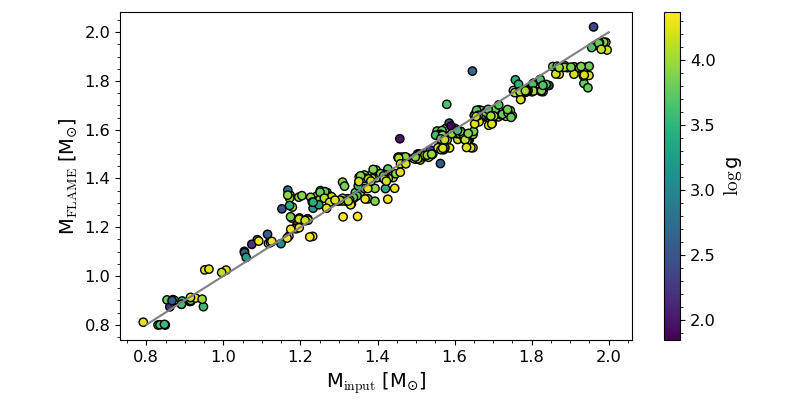}
    \includegraphics[width=0.4\linewidth]{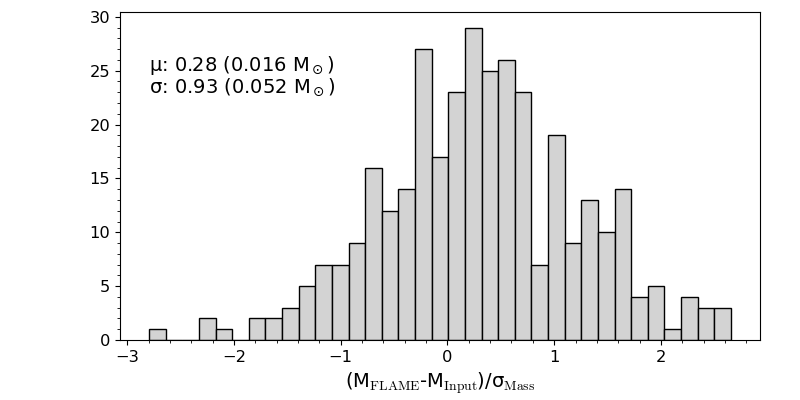}
    \includegraphics[width=0.43\linewidth]{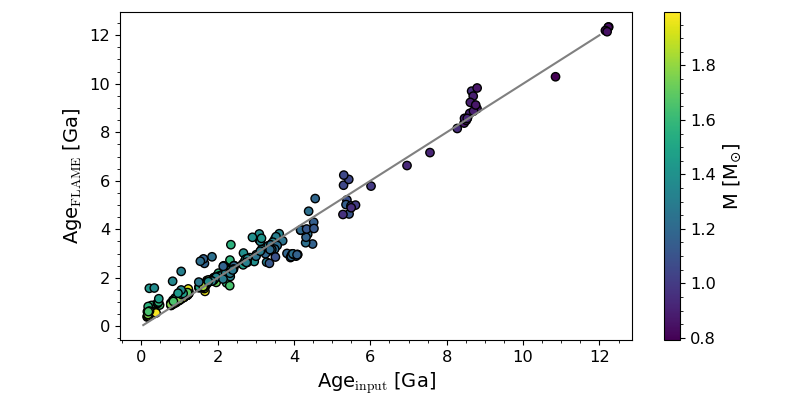}
    \includegraphics[width=0.4\linewidth]{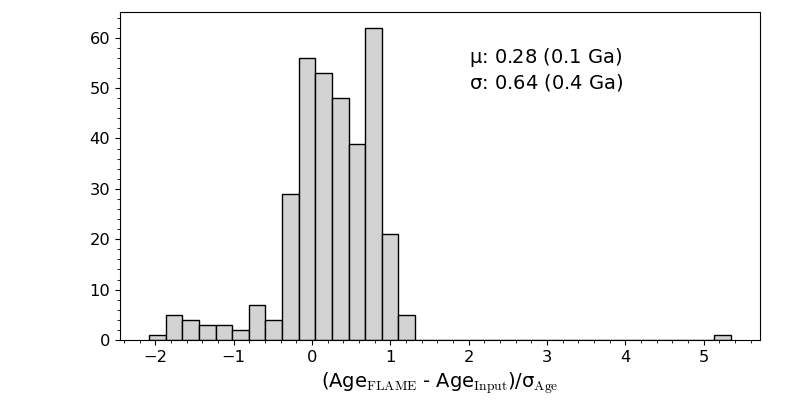}
    \caption{Comparison between the true input \mass\ (top) and age (lower) with the output from \flame.   The left panels illustrate a direct 1-1 comparison, colour-coded by \logg\ for mass and \mass\ for age, with the bisector in grey to guide the eye.   The right panels illustrate the histograms of the comparison of input-output scaled by their uncertainties.}
    \label{fig:simulateddataResults}
\end{figure*}

{The results of the analysis are shown in Fig.~\ref{fig:simulateddataResults}.  In the top panel we compare the input and retrieved masses, while the bottom panels compare the  ages.  The top left compares the input mass, \mass$_{\rm input}$, with \mass$_{\rm FLAME}$ colour-coded by \logg.   The identity line is also shown to help guide the eye and it can be seen that \flame\ retrieves the expected results, with no particular offsets or trends with other parameters.  In the right panel we show the histogram of the differences in mass but scaled by the \flame\ uncertainties.  As expected, the distribution peaks around zero, with roughly 68\% of the values lying within $\pm 1 \sigma$ ($0.016 \pm 0.052$ M$_\odot$).

The lower panel shows the same for the age, this time colour-coded by \mass.  Again no significant offsets or trends are visible, demonstrating that FLAME successfully retrieves the ages. The corresponding histogram shows the differences scaled by their uncertainties. Most of the values are within $\pm 1\sigma$, with an outlier at $-5\sigma$. Because the age parameter is highly non-linear, a Gaussian distribution is not expected; nevertheless, the histogram confirms that the ages are recovered to a satisfactory level (--$0.1 \pm 0.4$ Ga).}

\subsection{Validation using solar parameters and impact of observational errors\label{sec:valid-solar}}
The second set of tests used to validate our method focuses on the star we know best: the Sun. Our aim here is to assess the accuracy of both the models and the method, while also investigating the impact of each observational uncertainty on the resulting stellar parameters.

We simulated datasets representing the Sun by considering various scenarios of data quality.  For each scenario we make 500 simulated datasets.  
The best-case scenario is denoted as $\Sigma_0$, which assumes 1\% error in \lum, 60 K in \teff, 0.05 dex in \logg\ and \mh.    
We then reduce the precision in each of the observables one-by-one or in sets of two, as shown in the upper half of Table~\ref{tab:simulatedSunErrors}.   The left column is the name of the dataset and the subscript to $\Sigma$ indicates which parameter uncertainty is increased, e.g. $\Sigma_{LT}$ indicates that the uncertainties in both \lum\ and \teff\ were increased.

\begin{table}[]
    \caption{Impact of observational and systematic errors on the retrieval of the solar parameters. 
    }
    \label{tab:simulatedSunErrors}
    \centering
\resizebox{8cm}{!}{
    \begin{tabular}{lllllll}
         \hline
         \hline
         & $\sigma_L$ & $\sigma_{\teff}$ & $\sigma_{\logg}$ & $\sigma_{\mh}$ & $M\pm\sigma_M$ & $\age\pm\sigma_{\age}$\\
         & [\%] & [K] & [dex] & [dex] & [\msol] & [Ga]\\
         \hline
     $\Sigma_0$ & 1   &  60 & 0.05 & 0.05 & $0.99 \pm 0.05$ & $4.9 \pm 3.1$ \\
     $\Sigma_{L}$ & 3   &  60 & 0.05 & 0.05 & $0.99 \pm 0.04$ & $4.9 \pm 3.1$\\
     $\Sigma_{T}$& 1   &  120 & 0.05 & 0.05 & $0.99 \pm 0.04$ & $5.1 \pm 3.1$\\
     $\Sigma_{LT}$& 3   &  120 & 0.05 & 0.05 & $0.99 \pm 0.04$ & $5.1 \pm 3.1$ \\
     $\Sigma_{GM}$& 1   &  60 & 0.15 & 0.15 & $0.98 \pm 0.03$ & $5.9 \pm 4.3$  \\
     $\Sigma_{LTGM}$& 3   &  120 & 0.15 & 0.15 &$0.98 \pm 0.06$ & $6.1 \pm 4.3$ \\
     $\Sigma_{LTGM2}$ &10   &  250 & 0.25 & 0.25&$0.97 \pm 0.06$ & $6.0 \pm 4.4$  \\
     \\
     $\Sigma_{0,\rm Lp}$ & +5   &   &  &  & $1.01 \pm 0.04$ & $4.3 \pm 2.9$ \\
     $\Sigma_{0,\rm Lm}$ & --5   &   &  &  & $0.98 \pm 0.04$ & $5.7 \pm 3.2$ \\
     $\Sigma_{0,\rm Tp}$ &    &  +80 &  &  & $0.99 \pm 0.04$ & $5.0 \pm 3.1$ \\
     $\Sigma_{0,\rm Tm}$ &    &  --80 & &  & $0.99 \pm 0.05$ & $5.1 \pm 3.1$ \\
     $\Sigma_{0,\rm Mp}$ &    &   & +0.1 &  & $1.01 \pm 0.04$ & $4.8 \pm 3.1$ \\
     $\Sigma_{0,\rm Mm}$ &    &   & --0.1 &  & $0.97 \pm 0.04$ & $5.4 \pm 3.1$ \\
     $\Sigma_{0,\rm Gp}$ &    &   & & +0.1 &   $1.04 \pm 0.03$ & $1.7 \pm 1.7$ \\
     $\Sigma_{0,\rm Gm}$ &    &   & & --0.1 &   $0.93 \pm 0.03$ & $10.5 \pm 2.6$ \\
     $\Sigma_{LTGM,\rm Gp}$ &    &   & & +0.1 &   $1.04 \pm 0.03$ & $1.7 \pm 1.7$ \\ 
     $\Sigma_{LTGM,\rm Gm}$ &    &   & & --0.1 &   $0.96 \pm 0.06$ & $7.4 \pm 4.3$ \\

     \hline
         \hline
    \end{tabular}
}
\end{table}

The median \mass\ and \age\ of the set of 500 simulations, along with the median of their uncertainties are presented in Table~\ref{tab:simulatedSunErrors} in the last two columns.  We keep only a precision to two decimal places.  Here one can easily see for high precision measurements that the mass is retrieved accurately within the uncertainty.  The age is also retrieved correctly within the uncertainty. 
The largest impact happens when we decrease the precision in all measurements, i.e. from $\Sigma_0$ to $\Sigma_{LTGM}$ and $\Sigma_{LGTM2}$, and while the actual parameters do not change much (because there are no systematic errors), it can be seen that the uncertainties increase as expected.

\subsection{Impact of systematic errors on the solar parameters\label{sec:valid-sys}} 
We continue our sequence of validation tests by investigating how systematic errors in the observational data affect the inferred model properties.
We simulate datasets in the same manner as above by modifying one observational property at a time,
specifically 5\%, 80 K, 0.1 dex, 0.1 dex, for \lum, \teff, \logg, and \mh. 
We take as reference set $\Sigma_0$ knowing that the systematic errors will always be larger than the uncertainty.

The results of these tests are also given in Table~\ref{tab:simulatedSunErrors} in the lower half
of the table.  The subscripts L, T, G, M, denote \lum, \teff, \logg, and \mh, respectively, with 'p' and 'm' being a 'plus' or 'minus' systematic error by the amount specified in the table.  
As shown by the resulting parameters, the largest impact comes from the systematic errors in \logg\ 
where both \mass\ and \age\ no longer agree with the solar parameters.  The next parameter with most impact is \mh\ which changes primarily the mass.

To clarify if such a systematic error has the same impact in the case of worse precision parameters, we repeated the exercise using $\Sigma_{LTGM}$ for the systematic in \logg.  We see that with more freedom to explore the parameter space due to the lower precision, the systematic does not play such an important role.

This analysis here is restricted to some very specific test cases for a solar-like star.  
It is not possible to extrapolate the results to other parameter spaces or other combination of uncertainties and systematics.  For this, a full-scale analysis would be needed, and this is beyond the scope of this paper.


\section{Comparison with external datasets\label{sec:compare}}

In order to further validate our mass and age inference, we apply \flame\ to a set of existing catalogues and compare our results.
We begin with a comparison of \flame\ results with the Gaia Benchmark Stars \citep[GBS,][]{gbs-heiter15,gbsv3}.   We then compare the mass and age with results from catalogues that exploit asteroseismology.  Finally we illustrate a comparison with larger catalogue datasets using pre-operational DR4 data.

\subsection{Validation test with the Gaia Benchmark Stars (GBS) \label{sec:valid-gbs}}
The GBS are  
stars 
for which their \teff\ and \logg\ 
have been derived through fundamental relations, in particular using interferometry for measuring angular diameters.   The primary goal of the GBS is to serve as benchmarks for testing stellar physics in both atmospheres and interiors \citep[e.g.,][]{collet09, creevey15, creevey24}.  In addition they help calibrate stellar spectroscopic surveys, in particular Gaia, and allow one to derive high precision metallicities and abundances \citep{gbs-jofre14,gbs-jofre15,jofre-2017,casamiquela26}. They constitute a valuable reference for testing performances of pipelines, validating large-scale catalogues of stellar radii, masses, and ages, as well as for calibrating stellar system like globular clusters \citep[e.g.,][]{vandenberg14}.

{The most recent release of the Gaia Benchmark Stars, version 3 \citep[v3;][]{gbsv3}, comprises over 200 stars of which 190 have also been observed by Gaia and are present in \ac{gdr}3. 
In addition to the fundamental parameters \teff\ and \logg, the catalogue provides two sets of stellar masses, derived using different methodologies based on distinct stellar evolution models — the \basti\ and STAREVOL \citep{starevol1,starevol2} evolutionary tracks.  That paper focusses on the fundamental observed properties, and therefore stellar ages have not been published. }

As our stellar models extend only to the tip of the red giant branch (RGB), we selected approximately 140 stars that are unlikely to be red clump stars. The published values of {\lum, \teff, \logg, \feh}, together with their associated uncertainties, were then used as input parameters for processing with \flame.

{We first compare our derived radii with those from the reference study to verify consistency. Given that the methodology involves a degree of randomness, exact agreement is not expected. A comparison of the radii relative to their uncertainties yields a median difference of $-0.08\sigma$, while the median relative radius difference is $-0.002\sigma$, where $\sigma = 0.003$ \rsol. These results indicate that our derived radii are in high agreement with the published values. The good agreement is not surprising as we use similar stellar models and some of the same input data.  

{We compare our derived masses with their masses from the \basti\ evolution tracks ('MassB' from their catalogue) in Fig.~\ref{fig:comparegbs}.  In the top panel we show a mass-mass plot focussing on the region with $M < 2.2$ \msol\ (135 stars).  An identity line is also drawn to help guide the eye.  The colour code indicates the surface gravity and thus to a large extent the evolution status of the star.  Most of the MS stars (brighter colours) fall almost directly on the bisector, indicating strong agreement with their results.  A subset of the very evolved stars show an offset on the order of 0.1 -- 0.3 \msol, however, an insignificant offset when considering the uncertainties.  
The lower panel displays  a histogram of the mass differences scaled by the quadrature sum of their uncertainties, and it can be seen that most of the masses are within 2$\sigma$, indicating a very satisfactory agreement. 
{The few outliers at more than 4 sigma are for very evolved stars and these differences could be due to the fact that we use different input constraints - they use \teff, \feh, \lum, and \rad, and in this region of the HR diagram there is an important degeneracy between metallicity and mass.}
In Table~\ref{tab:gbsflameparams} (full version online) we list the \flame\ mass and age of the 140 GBS stars analysed here.}  

\begin{figure}
    \centering
    \includegraphics[width=0.85\linewidth]{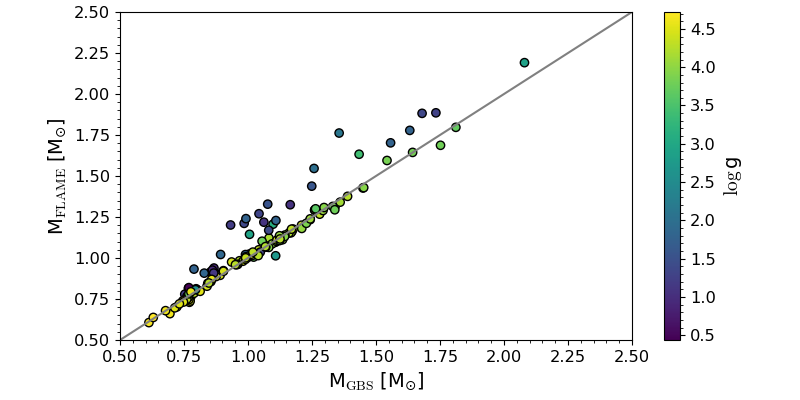}
    \includegraphics[width=0.80\linewidth]{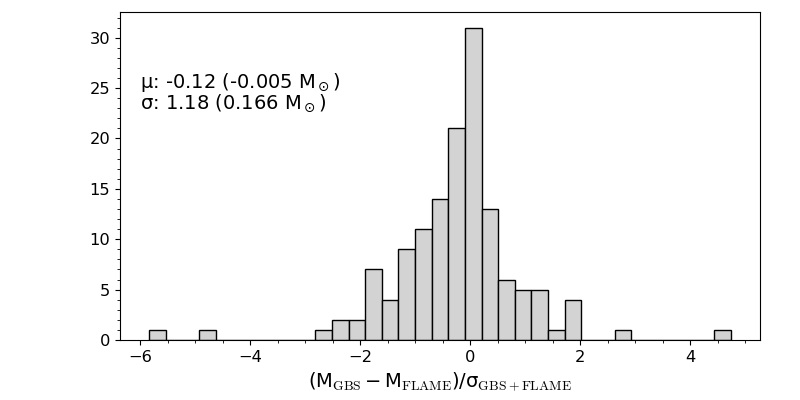}
    \caption{Comparison with GBS mass. }
    \label{fig:comparegbs}
\end{figure}

\subsection{Comparison with asteroseismic data}\label{sec:seismic}
Asteroseismology \citep[e.g.][]{ulrich97,aerts21} has the capacity to infer very accurate and precise stellar properties (surface gravity, radius, mass, and age, see e.g. \citet{lebreton2014, metcalfe14,hekker14,creevey17,mikkel25}  because the observed frequencies probe the inner structure of the star.   
Asteroseismic samples therefore provide an excellent dataset for which to validate our methodology.   We focus on two main dataset comparisons: a set of main sequence and sub-giant stars observed by both Kepler \citep{kepler} and the Apache Point Observatory Galactic Evolution Experiment \citep[APOGEE, ][]{apogee,apogee17}, as part of what is known as the APOKASC sample \citep{apokasc1}.  We use 
the results from \cite{serenelli17} who exploit 
the so-called global seismic quantities for a sample of approximately 450 stars.  The global seismic quantities are the mean large frequency separation $\langle \Delta \nu \rangle$ and the frequency of maximum power $\nu_{\rm max}$ which are sensitive to the global parameters of mass and radius.
Then we use the results from \cite{apokasc2} who provide results on a sample of approximately 6000 giants.   
For these two data set we perform a  new inference on the masses and ages and compare them with the seismic analysis.  

For the two data sets, we retrieve the \gspspec\ atmospheric parameters from Gaia DR3 and apply the published calibrations, along with the \flame\ luminosities. 
For the giants, we retrieve only the sources that correspond to a 'RGB' (red giant branch) status, because \flame\ (for GDR3 and GDR4) exploits models that span from the ZAMS to the tip of the red giant branch only.  
We only include those sources that have all four input parameters (\teff, \logg, \mh, \lum).

We illustrate the comparison between the resulting \flame\ mass and the asteroseismic mass on the left panels of Fig.~\ref{fig:compare_massage_seismic} by the background grey points.  Top and bottom panels correspond to the \citet{serenelli17}  and \citet{apokasc2}, respectively.
For the non-evolved stars, there is a relatively good agreement between the two masses with a median offset of --3\% but a large scatter giving a \ac{mad} of 11\%.  
For the giants, 
the median offset in relative mass is --7\% and the \ac{mad} is 13\%.  However, a linear trend with mass is also present, which gives an unsatisfactory agreement.  
This disagreement is not so surprising for stars in this evolution stage.  All stellar evolution tracks of giants of different masses and metallicities occupy a very narrow region in the HR diagram, with the result that even minor changes in the input parameters can change the inferred masses and ages significantly.   Similar trends were also seen for the mass and ages of giants published in Gaia DR3.  
In this comparison case, it is likely that the \flame\ results are incorrect, because we expect asteroseismology to well constrain the solution.  However, in the majority of comparisons with methods using similar input constraints as \flame\ it is not possible to know which results are the more accurate, and thus it is difficult to validate masses and ages of giant stars.

In order to fully validate \flame, we therefore make a new dataset comprising the same input \teff, \feh, and \logg\ reported in the asteroseismic catalogues.  For the \cite{apokasc2} catalogue, we use the parameter 'logg\_seis' which is the surface gravity derived from the asteroseismic analysis.  As it is usually very well determined, we add 0.05 dex in quadrature to the errors given in that paper for our test, in order to better reproduce typical uncertainties expected from atmospheric parameter analyses.  

In addition to these atmospheric parameters, we calculate new luminosities.   We exploit the Gaia DR3 G mag bolometric correction (BC) tool available through the Gaia cosmos webpages\footnote{\url{https://www.cosmos.esa.int/web/gaia/dr3-bolometric-correction-tool}} in order to derive new BCs using the input catalogue atmospheric parameters.  We apply the parallax bias correction to the Gaia DR3 parallax \citep{lindegren21}\footnote{\url{https://gitlab.com/icc-ub/public/gaiadr3_zeropoint}} and use the corrected parallax to estimate the distance. {For the \citet{serenelli17} sample, as no extinction value is available in their catalogue, and {\tt ag\_gspphot} is not available for the full sample, 
we exploit the extinction map from  \citet{vergely22} to derive extinctions for the full sample.    We then use the tool made available through Gaia\footnote{\url{https://www.cosmos.esa.int/web/gaia/edr3-extinction-law}} which is based on the work of \citet{danielsku2018} to convert \av\ to \ag.} 
For the APOKASC sample, we use their published value of extinction in the $V$ band $A_V$  and convert it to $A_G$ using an approximate conversion constant (the stars are all of similar \logg, \teff, and extinction so the extinction coefficient will be very similar for the subset).  
With these new datasets, we then re-run \flame\ for all of the stars in the two catalogues.

The resulting comparison of the masses is shown in Fig.~\ref{fig:compare_massage_seismic} on the left panels, but this time represented by the coloured points, with the colour-code indicating \logg.   It is immediately clear that the trend with mass has now statistically disappeared.  In addition the median offset between the relative masses is at 
{$<+0.01$\% with a \ac{mad} of 3\% for the non-evolved stars}, while it is 2\% with a \ac{mad} of 7\% for the giants, indicating very compatible results with the asteroseismic results.

Fig.~\ref{fig:compare_massage_seismic} right panels show  direct comparisons between the \flame\ age and the asteroseismic one.   The results are in agreement and follow the bisector.  The points deviating most from the bisector are also those points whose mass deviates most from the seismic mass.

\begin{figure*}
    \centering
    \includegraphics[width=0.49\linewidth]{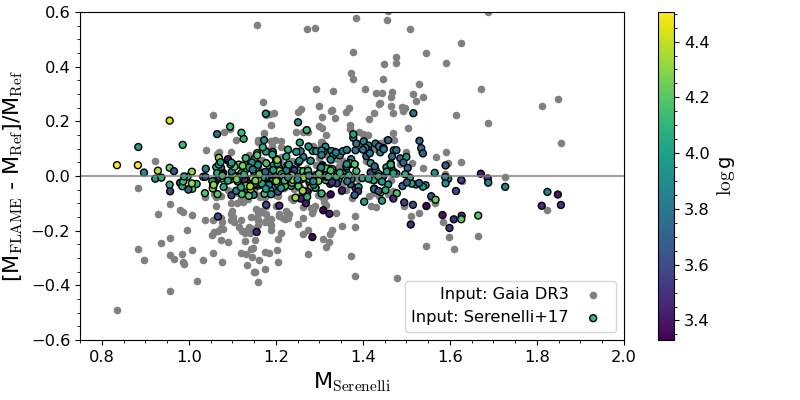}
    \includegraphics[width=0.49\linewidth]{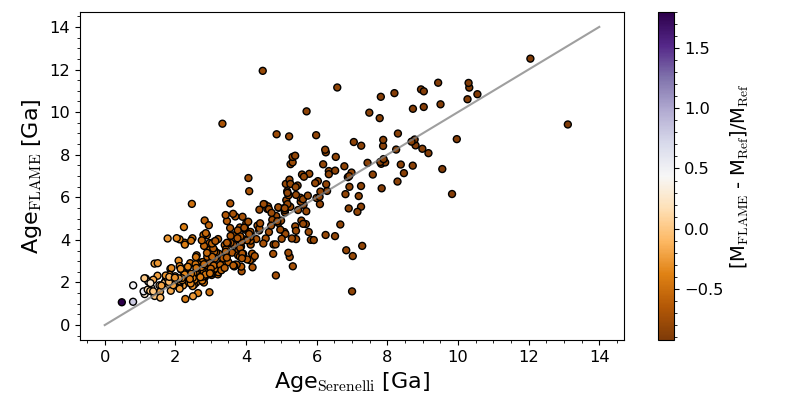}    
        \includegraphics[width=0.49\linewidth]{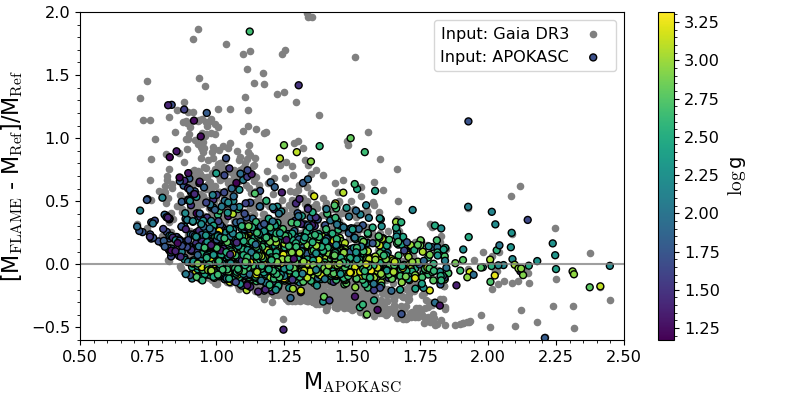}
\includegraphics[width=0.49\linewidth]{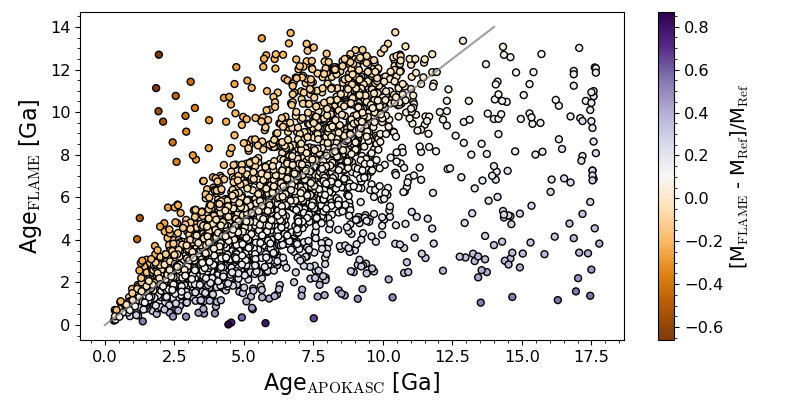}
    \caption{{\sl Left panels:} Relative difference between \flame\ mass and asteroseismic mass. The grey symbols show the results when we rerun \flame\ using GDR3 atmospheric parameters for these sources, while the colour-coded symbols show the comparison when we use the  atmospheric parameters from the {reference} catalogue (colour) as input to \flame.
    {\sl Right panels:} Comparison of \flame\ age and asteroseismic age when using the atmospheric parameters from the reference catalogue as input.}
    \label{fig:compare_massage_seismic}
\end{figure*}

\subsection{\flame\ DR4 performance}
In this subsection we present some results that are indicative of the DR4 performance of \flame. 
\subsubsection{StarHorse comparison \label{sec:starhorse}}
As a further validation test we exploit the StarHorse catalogue \citep{starhorse} which produces stellar parameters for around 10 million sources which are  in common between Gaia and other spectroscopic surveys.  They additionally derive stellar ages for approximately 2.5 million main-sequence and sub-giant stars by exploiting the PARSEC isochrones \citep{parsec12} and Gaia DR3 data.   This catalogue therefore provides a useful test set for \flame\ covering a large range of masses, ages, \teff, luminosities, and \mh.   
For this specific comparison test we use the Gaia DR4 \gspphot\ data as input and process the data using \flame.   As shown in the previous section, with different input data we expect to have different results.  This particular set then allows us to provide an overall comparison of how the \flame\ masses and ages in DR4 may compare with other catalogues.

Several test runs on Gaia DR4 data were performed during the development and pre-operational test phase.   The aim of these tests are to validate the scientific and technical parts of the processing and post-processing code.  They use as input a list of several million sources, comprising lists of known objects in catalogues that help to validate \flame\ and the other \apsis\ modules, and a list of randomly chosen sources to cover a large range of magnitude, parallax, sky coverage, and precision in these parameters.
A subset of the StarHorse catalogue is included in this test set.   Matching with the output from \flame\ we obtained a total of 381,254 stars in common all having \logg\ $>3.2$.

In order to show a meaningful comparison, we selected those stars with input data that were in broad agreement with the Gaia DR4 input data.  Specifically we required that the StarHorse \logg\ and the DR4 input \logg\ agreed within 0.5 dex,  the StarHorse \mh\ and the DR4 input \mh\ agreed to 0.5 dex, and the StarHorse \teff\ agreed with the DR4 \teff\ to within 400 K.   This resulted in a total of \num{344767} stars or 90\%  of the original sample.  
Fig.~\ref{fig:starhorse} shows the comparison between the two datasets for mass and age.

On the left panel we show the mass comparison for the 90\% sample.  There is good overall agreement; the median offset in mass over the full range is 3\% with the \flame\  masses being slightly higher, and the \ac{mad} is 5\%.  

In order to make a reasonable age comparison, we further restrict the sample so that the StarHorse mass matches the \flame\ mass to within  10\%.   
This results in a total of \num{285325} stars or 75\% of the original sample. 
We show the comparison of the ages for this subsample in the right panel of Fig.~\ref{fig:starhorse}, colour-coded by the StarHorse \logg.   
We can again see a very good agreement between the ages with a median offset of 
 --0.16 Ga and a \ac{mad} of 0.75 Ga. 
If we relax the constraint on the agreement in masses we find a mean offset of -0.1 Ga with a \ac{mad} of 1.0 Ga.

\begin{figure*}
\centering    \includegraphics[width=0.48\textwidth]{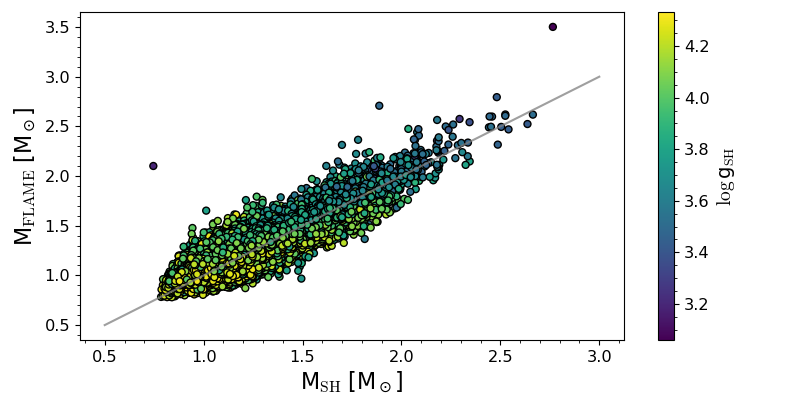}
\centering    \includegraphics[width=0.48\textwidth]{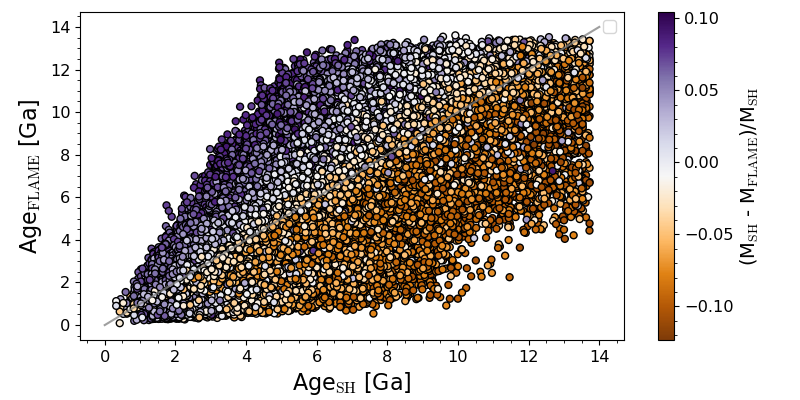}
    \caption{Comparison of the masses and ages from \flame\ using Gaia DR4 data as input and the results from the StarHorse catalogue.   
    {\sl Left:} Comparison of masses after imposing some constraints on the input data (see text, 344\,767 stars or 90\% of the original samples).  {\sl Right:} Comparison of ages after imposing the same constraints and restricting the stars to those with masses that agree to 10\% (269\,834 or 72\% of the original sample).
 \label{fig:starhorse}}
\end{figure*}

\subsubsection{Clusters \label{clusters}}
As a final validation test we investigate the distribution of ages from a few well known open clusters.  
We use the selection of cluster members from \cite{casamiquela2024}; and as input data we again use the pre-operational DR4 test run data as in Sect.~\ref{sec:starhorse}.  
We crossmatched the list of source IDs 
with the \flame\ dataset 
for the following clusters which represent a young, intermediate and older cluster:   
\object{NGC\,3532},  
\object{NGC\,2682},  
and \object{NGC\,188}.   

We show the distribution of the ages of each of the clusters in Fig.~\ref{fig:clusters} as the histogram, along with a kernel-density estimate (kde). 
The age of the cluster according to the literature is given in the label in each panel along with the median age value derived by \flame, and it can be seen that the 
\flame\ ages peak very close to the literature age, and only some outliers are seen.    The kde maxima only vary slightly from
the median age (6.2, 4.3, and 1.5 Ga, respectively).
For the youngest cluster, there is the largest dispersion.   
The outliers in each of the panels are often stars with low masses for which ages are difficult to derive, or stars for which the input parameters may not be in agreement with the expected values.   We remind readers that within the \apsis\ processing, all stars are processed on a star-by-star basis without any knowledge of it being associated with a cluster, so the input metallicities and extinction will be different for every star.

\begin{figure*}
    \centering
    \includegraphics[width=0.33\linewidth]{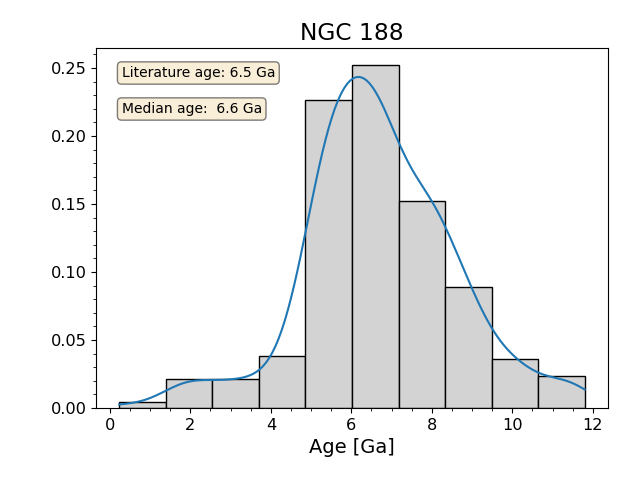}
    \includegraphics[width=0.33\linewidth]{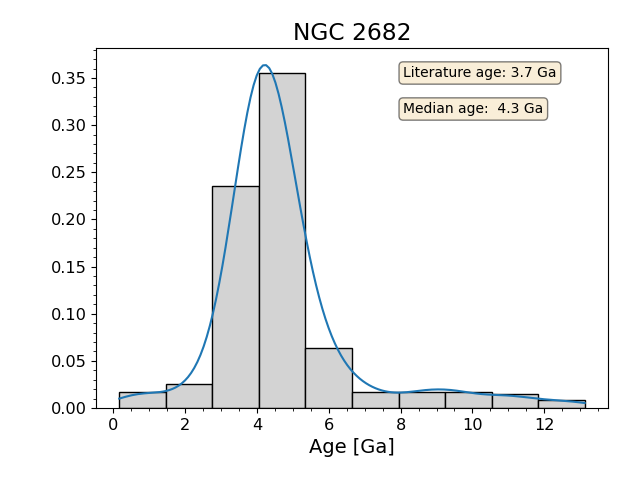}
    \includegraphics[width=0.33\linewidth]{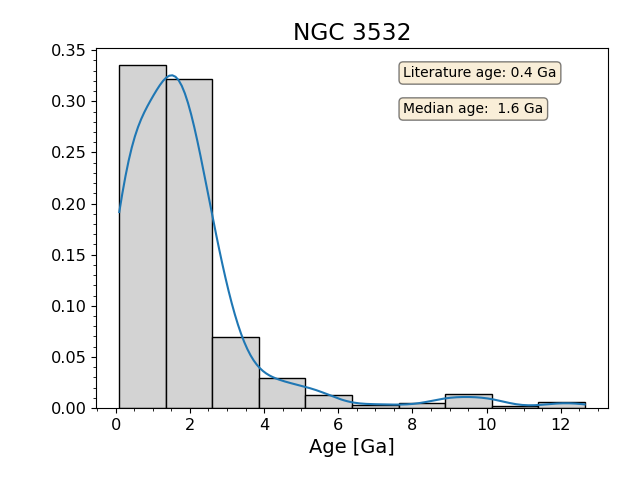}
    \caption{Distribution of ages for three clusters as derived by \flame\ using a dataset representative of Gaia DR4 data. }
    \label{fig:clusters}
\end{figure*}


\section{Application of \flame\ to datasets of high astrophysical interest \label{sec:apply}}
Having validated the methodology in Sect.~\ref{sec:validation}, and having compared results with some well established external data sets in Sect.~\ref{sec:compare}, we now use \flame\ to derive new masses and ages for various astrophysically interesting targets.

\subsection{High velocity stars\label{sec:exploit-hvs}}
High-velocity stars (HVSs, \citealt{katz25}) likely originate in different regions of the Galaxy and travel through diverse stellar environments. They are most commonly thought to be ejected by interactions with the supermassive black hole at the Galactic center, though alternatives such as stellar explosions or cluster dynamics have been proposed. Determining their masses and ages can provide key insights into their origins.
With this in mind, we aim to derive the ages of the  the sample of high velocity and metal-poor stars presented in \citet{katz25}.

As input data to \flame\ we use the \teff, \logg, \feh, and \alphafe\ as given in \citealt{katz25}.  As there are no observational errors published with the  \teff\ and \logg, we adopt a classical 100 K and 0.1 dex as an error for all of the stars. 
We use Eq.~\ref{eqn:alphacorrection2} to calculate  \mh\ from \feh\ and \alphafe, and we adopt the published error on \feh\ as the error on \mh.  
We use these spectroscopic parameters to derive the BC.

In order to derive the luminosities, we extract complementary data from Gaia DR3; we use the astrometric information and derive the parallax bias to correct the published parallaxes.  
In addition, as only 35 of the 51 stars have published extinction values, {we follow the 
procedure mentioned in Sect.~\ref{sec:seismic} to derive new extinction values.}
The resulting extinction values are broadly in agreement with the existing \ag\ in Gaia DR3.

Using the corrected parallax, BC, \ag, and Gaia DR3 $G$, we calculate the luminosities using Eqs.~\ref{eqn:def-luminosity} and \ref{eqn:def-magbol}.
The input data set to \flame\ therefore comprises \{\lum, \teff, \mh, \logg\}, and we derive their masses and ages 
for the full sample.  

We show the distributions of the ages of the high velocity sample in Fig.~\ref{fig:hvs_ages} and we publish their luminosities, radii, masses, and ages in Table~\ref{tab:hvs_agemass}.   
The distribution of the ages peak around 11 Ga, corresponding to the era of the Halo phase of the Milky Way.  
It is not unexpected to find some younger stars in the sample because we do not know the exact origin of their star formation.

\begin{figure}
    \centering
    \includegraphics[width=0.98\linewidth]{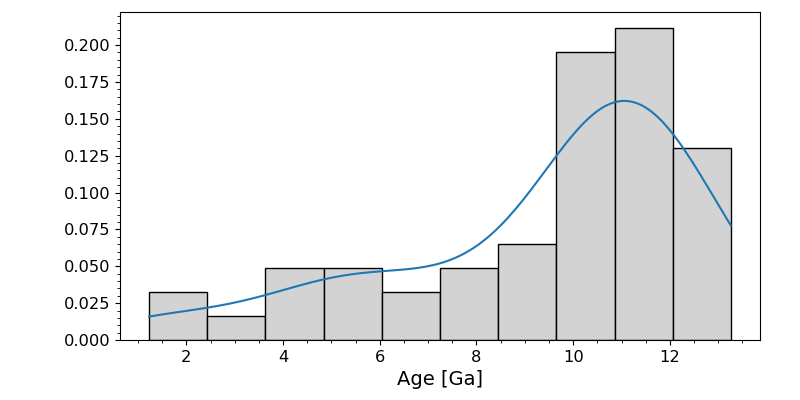}
    \caption{Distribution of ages derived with \flame\ of high velocity stars from \citet{katz25}.}
    \label{fig:hvs_ages}
\end{figure}

\subsection{Solar-like stars with very low-mass companions \label{explot-lowmass}}

Low-mass stars burn nuclear fuel extremely slowly, and can live for tens to hundreds of billions of years.  Their observable surface properties evolve only minimally over time, and therefore their masses and ages remain poorly constrained.  In particular, if we consider objects on the boundary of sub-stellar and stellar objects, knowing their masses and ages can lead to a better understanding of their formation and evolution.  

 When such objects are members of binary systems with more massive primaries, the primary stars can often be characterized much more accurately using independent techniques such as spectroscopy, stellar modeling, and activity or rotation diagnostics. By deriving reliable ages and masses for these primary stars, we can indirectly constrain the physical properties of their low-mass companions. 

 With this in mind, we consider the sample of 36 stars with low-mass companions studied by \citet{maxted2025}; of which eight are in fact transiting exoplanets.  
 The low-mass companions contribute less than 0.1\% of the total flux in the V band.  This means that we can use assumptions of an isolated star to make a meaningful determination of the properties of the primary component.

 To construct our data set of \{\lum, \teff, \logg, \feh \} we follow the same approach as described in Sect.~\ref{sec:exploit-hvs}.  We adopt the published \teff, \logg, and \feh\ from \citet{maxted2025}, and we calculate the primary
 star's luminosity using $G$, \ag, BC, and a corrected parallax (Eqs.\ref{eqn:def-luminosity} and \ref{eqn:def-magbol}).   We then process this dataset with \flame. 
 
 The resulting masses and ages of these systems are given in Table~\ref{tab:lowmass_agemass} where we use the published mass ratio ($q$) as given by \cite{maxted2025} to calculate the companion masses.   
 In Fig.~\ref{fig:lowmass_agemass} we also illustrate 
 the mass of the companion star as a function of the stellar age.
{We note that $q$ has a very small dependence on the the mass estimate, but as our mass estimates are in close agreement with those of \citet{maxted2025} (2\% $\pm$ 3\%), $q$ should not change significantly.}

 \begin{figure}
     \centering
     \includegraphics[width=0.98\linewidth]{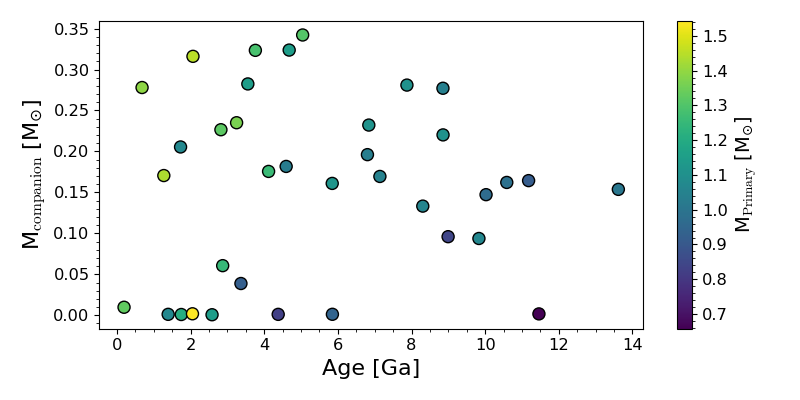}
     \caption{
     Companion mass as a function of stellar age for the sample of 
     systems with low-mass companions in \citet{maxted2025}.
     }
     \label{fig:lowmass_agemass}
 \end{figure}

\subsection{PIC}
PLATO \citet{plato2025} is an ESA mission currently planned to launch in January 2027.   Its objective is to search for and characterise exoplanets, in particular those like the Earth.  Unlike Gaia which scans the full sky, PLATO  will stare at a fixed field of view for at least two full years, allowing the collection of very long and high precision datasets capable of detecting Earth-like planets around Sun-like stars.   As part of the preparation of the PLATO data validation, a selection of stars known as the science, calibration, and validation (scv) PLATO Input Data (PIC) stars has been made to help validate the PLATO data and pipelines.  
A subset of the scvPIC has been prepared by the PLATO Benchmark Stars work package \citep{merle26} and is presented in \citet{zwintz26}. 

About 200 of the targets in the scvPIC from \citet{merle26} comprise high quality samples of data from Gaia DR3 known as the golden sample of astrophysical parameters \citep{creevey23}, members of open clusters and targets that can be measured with interferometry.  We refer readers to \citet{zwintz26} for details on the target selection and the description of the targets that made it to the scvPIC.  
In this section we aim to provide new masses and ages of this dataset by exploiting the astrophysical parameters from Gaia DR3, as produced by \gspphot\ and \flame.   We use as input data the \lum, \teff, \mh, and \logg, as given directly in the Gaia archive, and we process these datasets with the Gaia DR4 version of \flame.  

The new masses and ages of these stars are given in Table~\ref{tab:pic}, and in Fig.~\ref{fig:pic} we illustrate the HR diagram colour-coded by the resulting age.   These homogenous parameters could be used for validating some of the first stellar parameters from PLATO.  

\begin{figure}
    \centering
    \includegraphics[width=0.95\linewidth]{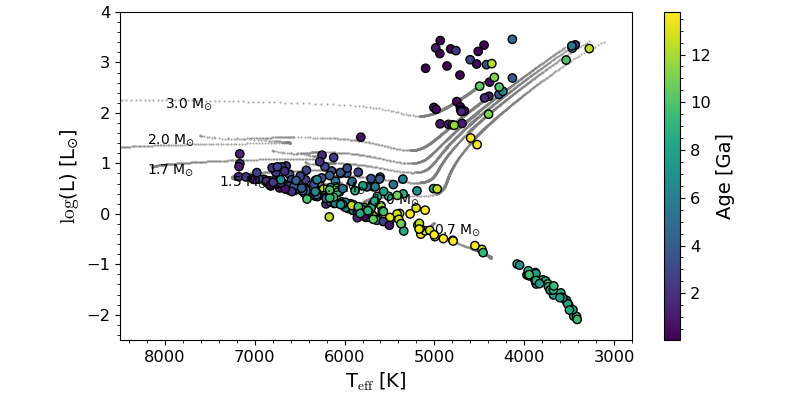}
    \caption{HR Diagram of a subset of the benchmark scvPIC stars from the GDR3 Golden Sample colour-coded by age.} 
    \label{fig:pic}
\end{figure}

%
\section{Specifics of Gaia Data\label{sec:specs}}
Sects.~\ref{sec:validation} and \ref{sec:compare} focussed primarily on the validation of the model inference method \algotwo, while Sect.~\ref{sec:apply} applied the model-based inference to a set of sources of astrophysical interest.
In this section, we revert to the complete methodology and examine the expected performance of \flame\ in the context of Gaia data.

\subsection{Inflation of input errors from \gspphot\ and \gspspec \label{ssec:errorinflation}}

As observed in GDR3, the formal uncertainties in atmospheric parameters can be  typically on the order of 10 K in \teff\ and less than 0.03 dex in \logg\ and \mh.  
As explained in \cite{gspphot} and \cite{gspspec} these formal uncertainties reflect the SNR of the data and even if systematic errors may be present, these can not be captured by the uncertainties in the fitting method.   \cite{gspphot} explain that by comparing with external data, typical errors are on the order of 110 K for \teff, and 0.2 -- 0.25 for \logg.
While it is essential to report formal uncertainties for the atmospheric parameters that are derived directly from the data, these small uncertainties can have a negative impact in particular on the model-based \flame\ parameters.  Firstly, the uncertainty in the bolometric correction is often on the order of 0.001 mag, which means it contributes minimally to the total error budget for luminosity, leading to unrealistically small uncertainties. Secondly, in model inference, if the uncertainties are overly constrained, the resulting mass and age estimates from \flame\ may be biased, as discussed in Section 4.3 for the solar case, and will likely have unrealistic error margins.
To address this, we have introduced an inflation parameter for the atmospheric parameter uncertainties in GDR4.

For \gspspec\ data, this has been introduced in a straight-forward manner by adding in quadrature 60 K, 0.1 dex, and 0.1 dex in the \teff, \logg\ and \mh\ parameters respectively, before the BC is evaluated.  This implies that the inflated uncertainties get propagated all the way to the mass and age inference.  

For GSP-Phot, the approach differs, as \flame\ does not process the published values along with their confidence intervals. Instead, it directly processes the MCMC samples.   As the MCMC samples retain correlations among the independent parameters, the inflation has to be done by maintaining these correlations.  In practice this implies the multiplication of the distance of each sample point from its mean by a relative factor $T/\teff$.  It, in effect, stretches the samples away from each other and maintains the correlations. 
The value of $T$ has been determined empirically by using multiple data samples and comparing their mass and age parameters with those from \flame\ after scaling by the new \flame\ uncertainty. In the ideal case, the agreement between the datasets falls within $\pm 3\sigma$ for almost all of the targets. The chosen value of $T$ corresponds to 50 K.  
The MCMC samples have five independent fitting parameters: \{\teff, \mh, \logg, \ag, $d$\}.   Thus, the inflation of the parameters has been applied by accounting for all the correlations within the data.

\begin{figure}
    \centering
    \includegraphics[width=0.9\linewidth]{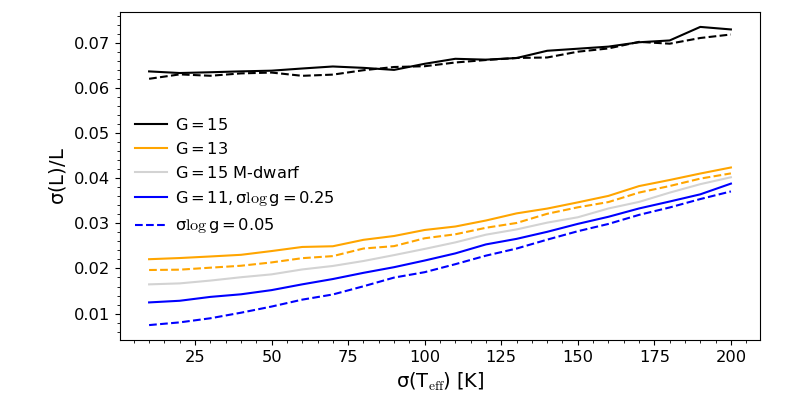}
    \caption{Luminosity errors as a function of input \teff\ uncertainty for different $G$.}
    \label{fig:properror}
\end{figure}

\begin{figure}
    \centering
    \includegraphics[width=0.95\linewidth]{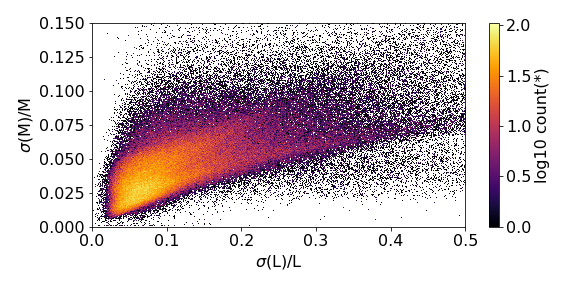}
    \caption{Mass errors as a function of luminosity errors for a sample of approximately 10 million stars.  
    }
    \label{fig:properrormass}
\end{figure}
\subsection{Expected uncertainties\label{ssec:uncert}}

\subsubsection{Luminosity\label{ssec:lumerrors}}

{The main purpose of \algoone\ is to derive the stellar luminosity.  
It depends on numerous observations, cf. Eqs.~\ref{eqn:def-luminosity}
and \ref{eqn:def-magbol}, namely \{\teff, \logg, \mh, $f_G$, $\varpi$\}.
The combination of the atmospheric parameter uncertainties are responsible for the  uncertainty in BC, and then this along with $f_G$ and $\varpi$ determine $\sigma(L)$.  
The uncertainties $\sigma(f_G)$ are closely tied to the stellar magnitude, while $\sigma(\varpi)$ depends not only on magnitude, but also the spectral types and luminosity classes. For example, for solar-like stars the relative parallax error will increase with $G$, while for hot $A$ stars we will also see a relative increase, but of a different order compared to the solar-like star case. 
In addition, the SNR of the input spectra from which the APs are derived also depend on $G$: the brighter the star, the higher the SNR, the smaller the uncertainties in \{\teff, \logg, \mh\}.  However, the performance of the atmospheric parameter modules will also vary from star type to star type.}

To show the performance of the determination of \lum, we consider a solar-like star, i.e, $|(L/L_\odot - 1)| < 0.15$ and {\tt spectraltype\_esphs} = 'G' and search for a sample of these stars at $G = 9, 11, 13, 15, 17$ in GDR3.  We use the mean value of the parameters {\tt g\_mean\_flux\_over\_error}, {\tt varpi}, and {\tt varpi\_error} as associated representative uncertainties at each of the $G$ magnitudes.  
At the same time we also consider uncertainties on \teff\ that vary between 10 and 200 K, and $\sigma_{\logg} = \sigma_{\mh}$ which vary between 0.025 and 0.250 dex.

For each \teff\ and \logg\ error, we performed bootstrap simulations to derive the BC, and using the set of expected values for the astrometry and photometry, along with the Eqs.~\ref{eqn:def-luminosity}
and \ref{eqn:def-magbol} we then calculate \lum.   We use the standard deviation of \lum\ as an estimate of the expected uncertainty, and then calculate the relative uncertainty, $\sigma(L)/L$.  

This is illustrated in Fig.~\ref{fig:properror} where we show by the blue dashed lines $\sigma(L)/L$ as a function of $\sigma(\teff)$ assuming $\sigma_{\logg} = \sigma{\mh} = 0.05$ dex for a solar-like star with $G=11$.  As $\sigma(\teff)$ increases, $\sigma(L)/L$ varies from about 0.5\% to 3\%.   The continuous line represents the results when we increase the $\sigma_{\logg}$ to 0.25 dex.
As can be noted, the uncertainty in \logg\ and \mh\ also play a small role in determining $\sigma(L)/L$.

The orange lines show results for an assumed solar-like star at $G=13$.  The difference between the impact of $\sigma_{\logg}$ going from 0.025 (dashed lines) to 0.250 dex (continuous lines) is much smaller than for the $G=11$ case.  Here, the parallax and photometry begin to have a more important role in $\sigma(L)/L$, and increase the minimum uncertainty to 2\%.
The black lines represent $G=15$ and we can now see that there is virtually no impact due to varying $\sigma_{\logg}$ (no difference in dashed and continuous lines) while $\sigma_{\teff}$ still plays a small role by increasing $\sigma(L)/L$ from just over 6\% to just over 7\%.  
Doing the same exercise for a $G=17$ star, we are entirely dominated by the parallax and photometry errors and $\sigma(L)/L$ remains at a constant 35\% (not shown).

The results above are representative of predicted uncertainties for a solar-like star, using GDR3-like astrometry and photometric precision. 
To illustrate the predicted uncertainties for an M-dwarf for $G=15$, we follow the same methodology and illustrate only the result for $\sigma_{\logg} = 0.25$ by the lightgrey solid line.  Here $\sigma(L)/L$ varies from 1.5\% to 4\%.  In this regime the parallaxes are quite large (these are nearby stars) and the parallax uncertainties therefore play a much smaller  role.
  We expect all of these values to improve in GDR4 due to the expected increase in precision of the astrometric and photometric data.

\subsubsection{Radius\label{ssec:radiuserrors}}

As \rad\ is derived directly  from \lum\ and \teff, through the Stefan-Boltzmann law, the uncertainties are straight-forward to predict from a standard propagation of errors:
\begin{equation}
    \sigma(R) = \sqrt{ \frac{\sigma^2_L}{4\lum} + \frac{4\sigma^2_x}{x^6}},
    \label{eqn:properrorSB}
\end{equation}
where $x = \teff/T_{\odot}$.

By considering four of the limiting cases from the above shown in Fig.~\ref{fig:properror} we calculate for $G=11$ with $\sigma_{\teff}$ = 50 K and 200 K, $\sigma(R) = $ 0.02 and 0.07 \rsol, while for $G=15$ it varies from 0.035 to 0.08 \rsol, where here we have considered that the star is 1\rsol.

\subsubsection{Mass \label{ssec:masserrors}}
If we were to use a relation such as the $M-L$ relation for main sequence stars,  e.g. with index $\sim$3, we could approximate $\sigma(M) \sim \sqrt{\frac{\sigma^2_L}{4L}}$.  For the cases illustrated above, where $\sigma_L$ ranging from 0.014 to 0.70 \lsol, $\sigma(M)$ would vary between 0.007 and 0.035 \msol.
However, in \flame\ the mass is constrained by a set of physical values set by the stellar evolution models.
For a single metallicity, the mass (and age) are constrained additionally by \teff\ and \logg.
On top of that, the sensitivity of each of the observations vary with mass and evolution stage.  
To illustrate the typical uncertainties in mass that we obtain from \flame, we show in Fig.~\ref{fig:properrormass} the results from one of pre-operational test runs, described earlier in Sect.~\ref{sec:starhorse}, on approximately 10 million stars spanning a wide range of masses, evolution stages and metallicities with over 90\% having $G<17$.  
In order to make a comparable plot, we show how the relative uncertainty in \mass\ varies with relative uncertainty in \lum, with the colour code indicating the number of sources.  Yellow indicates a bin (point-like) with approximately 100 stars while black indicates 1 source.   The majority of the relative uncertainties from \flame\ are typically less than 20--30\% in \lum, while less than 10\% in mass.    
We indeed observe similar $\sigma_M/M$ to that from a $M-L$ relation, where typical 7\% uncertainties give between 2 -- 4 \% $\sigma_M/M$.

\subsubsection{Age \label{ssec:ageerrors}}
Age uncertainties have a highly non linear dependence on the uncertainties in the other parameters. The uncertainty depends not only on the observational errors but also on the star’s evolutionary stage and mass. Moreover, when the inferred age is near the boundaries of the model grid —whether very young or very old— the uncertainty limits are truncated at the upper or lower ends. To illustrate the expected performance of the age estimates, we use the same sample of approximately 10 million stars mentioned above and examine the age uncertainties within specific mass ranges.
Additionally we focus on MS stars and restrict the metallicity to $\mh > -0.5$, 
as metallicity significantly impacts a star’s evolution and, consequently, its age.

Fig.~\ref{fig:age_errors} illustrates the distribution of age uncertainties for mass ranges spanning the FGK range (different colours).  The top panel shows normalised histograms of relative age uncertainty while the lower figure shows normalised histograms of the age error in Ga.   As this figure illustrates, the lower mass stars tend to have very large relative uncertainties on the order of 70\% which decreases slightly with mass.  This is due to the relatively slow rate of change of the surface properties of these low mass stars during the MS, and thus the incapability of the observed properties to well constrain the age (given their unknown masses).  
The higher mass stars (red) will have relative age uncertainties on the order of 20\% because  \teff, \lum, and \logg\ change relatively quickly as the star ages, and thus these properties allow to narrow in efficiently on their ages.
For solar-like stars (green) we have a broad distribution of uncertainties spanning 15\% -- 60\%, that depends not only on the input observational errors but the actual assigned age which statistically covers the full range of age.
One can also note that despite the attempts to inflate the uncertainties, there are still some stars where the relative uncertainties are on the order of less than 5\%, and this happens for approximately 1\% of the 10 million sample.  

If we generate the same figures for all stars beyond the main sequence (i.e. evolution stage $>420$), the distributions change significantly. For lower-mass stars ($\sim 0.8$ \msol), the uncertainties peak around 10\% and increase to 15–20\% for higher masses.

\begin{figure}
    \centering
    \includegraphics[width=0.95\linewidth]{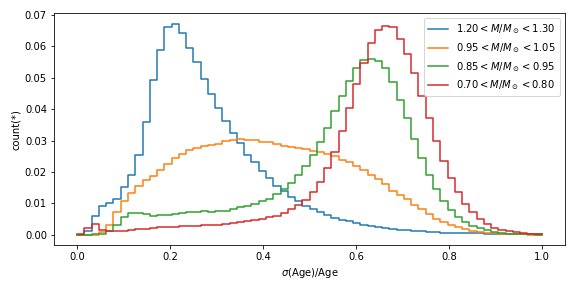}
    \includegraphics[width=0.95\linewidth]{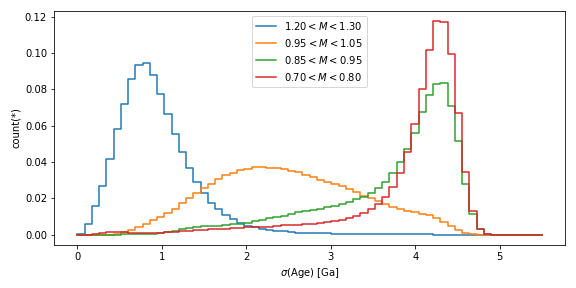}
    \caption{Histograms illustrating the distributions of relative age uncertainty (top) and absolute age uncertainty (lower) for MS solar-metallicity stars in different mass ranges. }
    \label{fig:age_errors}
\end{figure}



\section{Conclusions\label{sec:conclude}}
The FLAME pipeline is one of the official software tools used within the Gaia Data Processing and Analysis Consortium to produce astrophysical parameters of sources that appear in the Gaia Data Releases.  Specifically, it is designed to estimate luminosity, radius, radial velocity correction due to gravitational redshift, mass, age and  evolutionary status.  
In this paper we laid out the architecture of the code, explained the methodology and models and described the input Gaia data that is processed by the code.
The code is separated into an analytical part and a model-inference part which allows all the complexities of the model atmospheres and passbands to be encapsulated into a single value in the former -- the bolometric correction.   The model inference to estimate mass, age, and evolutionary status,  then compares the physical intrinsic properties of the star with those predicted from the BaSTI stellar evolution tracks, namely, the luminosity, metallicity, surface gravity and effective temperature.

We then focussed our efforts on validating the model-based inference part of the code to estimate the mass and age.  We did this by testing \flame\ on simulated stars and the Sun in Sect.~\ref{sec:validation}, and then by comparing our results with those of external catalogues in Sect.~\ref{sec:compare}.  
We found that when we use the same input data as the external catalogue we obtain masses and ages that are in tight agreement with those of the external catalogues.  However, when we use a different source of input data to the external catalogue, our results broadly agree in mass and age across main sequence and sub-giant stars, but the results for giants are not comparable.  This is due to the fact that the evoluted models all occupy a very tight region in the HR diagram during this phase and therefore any small change in input data can lead to a very different optimal model.

In Sect.~\ref{sec:apply} we applied \flame\ to several sets of stars in order to produce new luminosities, masses and ages:  a high radial velocity metal-poor sample from \cite{katz25}; a sample of stars with very low mass companions from \cite{maxted2025}; and finally a sample of high quality bright stars from the PLATO field of view.  Their parameters are given in the appendix and made available online.

We discussed specifics of \flame\ related to the Gaia DR4 data, such as the addition of a metallicity offset to ensure consistency between the input and model  metallicity (Sect.~\ref{sec:method}), the source of the atmospheric parameters (Sect.~\ref{sec:data}), the inflation of uncertainties and the expected uncertainties in Gaia DR4 (Sect.~\ref{sec:specs}).  
Uncertainties in luminosity depend not only on the $G $ magnitude, parallax, and parallax uncertainty, but also on the spectral type and the quality of the input atmospheric parameters.  The precision on these latter plays a very different role  depending on the spectral type and  magnitude of the star (which together intrinsically relate to the parallax).  They can reach as low as 1\% for a $G=11$ star $G$-type star (luminosity close to 1) even when errors on the \teff\ are around 50 K.   For the same type of star with a magnitude of $G=15$, the quality of the atmospheric parameters plays practically no role, showing a flat distribution around 6--7\%.  However, a $G=15$ M-dwarf star can reach 1--2\% uncertainties in luminosity. 
Uncertainties on the radius depend entirely on the precision in luminosity and \teff\ because the Stefan-Boltzman law is invoked.  For Gaia data, the radius uncertainty can reach as low as 1-2\%. 
Mass uncertainties follow roughly the expected uncertainty as predicted by an analytical mass-luminosity relation, but stellar models, help to constrain the uncertainties better, and tests during the pre-operational development phase for Gaia DR4 show typical mass uncertainties peak between 2\% and 6\%.
For FGK stars we expect to retrieve ages with no better than 50\% for main sequence stars with masses below 0.8 \msol.  For masses between 1.2 and 1.3~\msol, expected precisions peak around 20\%.    For giants we find the opposite trend, with typical precisions on the order of 10\% for stars with masses around 0.8~\msol, and the uncertainties increasing for higher masses.

\flame\ is  a powerful and efficient tool for deriving the masses and ages of stars from Gaia data. The results demonstrate that the derived masses are expected to be accurate to within 5\%, while the ages can be estimated with a reliability of as good as 10\%, provided the input data is of high quality. These findings underline the potential of FLAME to contribute significantly to the field of galactic archaeology with data from Gaia DR4, offering valuable insights into the history and evolution of our Galaxy, and to the characterisation of exoplanetary systems.

\begin{acknowledgements}
This work has made use of data from the European Space Agency (ESA) mission
{\it Gaia} (\url{https://www.cosmos.esa.int/gaia}), processed by the {\it Gaia}
Data Processing and Analysis Consortium (DPAC,
\url{https://www.cosmos.esa.int/web/gaia/dpac/consortium}). Funding for the DPAC
has been provided by national institutions, in particular the institutions
participating in the {\it Gaia} Multilateral Agreement.

SC acknowledges financial support from PRIN-MIUR-22: CHRONOS: adjusting the clock(s) to unveil the CHRONO-chemo-dynamical Structure of the Galaxy” (PI: S. Cassisi) finanziato dall’Unione Europea – Next Generation EU, and Theory grant INAF 2023 (PI: S. Cassisi). 
OK acknowledges support by the Swedish Research Council (grant agreement no. 2023-03667) and by the Swedish
National Space Agency.
CN acknowledges financial support from the Centre national d’études spatiales (CNES), France (ROR: https://ror.org/04h1h0y33) within the framework of the Gaia mission.

\end{acknowledgements}

   \bibliographystyle{aa} 
   \bibliography{flame} 

\appendix

\section{Author contributions}
The authors of this paper have contributed in many ways to the \flame\ software and feedback on performance.  The \flame\ team comprises OLC (scientific lead), CO (main java developer), FT, YL, LC, BP, and have contributed significantly to all aspects mentioned below.   Significant support in the development of its integration into the CNES framework and  scientific validation framework has been provided by NBa, FP, CR, CBJ, MF, RA, and AB.  
BE, OK, SC, AK, RS have contributed to the provision of input models, bolometric corrections, and simulations.  DR has provided software support for the python-java development of the \spins\ code.   
NBr, AL, GK, CS, RS, AK, CN, RA, MF, CB have contributed to the scientific validation and feedback look. 
All authors have contributed to writing this manuscript.

\section{\flame\ stellar parameters of the GBS}

\begin{table}[h]
    \caption{New masses and ages of a subset of the GBS. }
    \label{tab:gbsflameparams}
    \centering
   \begin{tabular}{lccccc}
    \hline \hline
    Gaia DR3 source ID  & $M_{\odot}$ & Age \\
     &   [M$_\odot$] & [Ga]\\
\hline
6904703228801028864&   1.096 $\pm$   0.015&   7.8 $\pm$   0.6 \\
 2195115561168483712&   0.987 $\pm$   0.038&  11.0 $\pm$   1.6 \\
 2190891581091291520&   1.201 $\pm$   0.044&   4.7 $\pm$   0.6 \\
 1872046609345556480&   0.661 $\pm$   0.009&  10.7 $\pm$   2.3 \\
 1872046574983497216&   0.606 $\pm$   0.011&   7.8 $\pm$   4.0 \\
 1787990525934491776&   0.754 $\pm$   0.007&  13.0 $\pm$   0.7 \\
 2278918344568877056&   1.595 $\pm$   0.060&   1.8 $\pm$   0.1 \\
 6412595290592307840&   0.702 $\pm$   0.008&  12.1 $\pm$   1.6 \\
 1891598193816300544&   1.234 $\pm$   0.020&   3.1 $\pm$   0.8 \\
 1907131544341497600&   4.097 $\pm$   0.168&   0.2 $\pm$   0.1 \\
 1982252790090609152&   1.328 $\pm$   0.038&   3.7 $\pm$   0.3 \\
 1876331990259358976&   4.143 $\pm$   0.134&   0.2 $\pm$   0.0 \\
 2719475542667772416&   1.163 $\pm$   0.016&   4.7 $\pm$   0.3 \\

    \hline\hline
    \end{tabular}
        \tablefoot{The full table of 140 stars is available online.}

\end{table}

\section{Mass and age of high velocity stars}

\begin{table*}[h]
    \caption{Newly derived luminosities, masses and ages of the high velocity star sample from \cite{katz25}. }
    \label{tab:hvs_agemass}
    \centering
    \begin{tabular}{llcccc}
    \hline \hline
    Star & Gaia DR3 source ID & [M/H] & $L_\star$ & $M_{\odot}$ & 
    Age \\
     & & & [L$_\odot$] &  [M$_\odot$] & [Ga]\\
    
    \hline
Gaia\_2 & 1336408284224866432&  -1.783&   2.145 $\pm$   0.036&   0.807 $\pm$   0.054&  10.4 $\pm$   2.9 \\
 Gaia\_108 & 139417803820690048&  -0.366&   1.743 $\pm$   0.029&   0.959 $\pm$   0.075&   8.8 $\pm$   3.4 \\
 Gaia\_45 & 1839706467664901248&  -1.126& 124.907 $\pm$   2.108&   0.839 $\pm$   0.032&  12.1 $\pm$   1.2 \\
 Gaia\_41 & 2726872575862585856&  -1.634&   4.396 $\pm$   0.074&   0.836 $\pm$   0.038&  10.8 $\pm$   1.3 \\
 Gaia\_56 & 2618198324041826432&  -1.926&   1.162 $\pm$   0.020&   0.764 $\pm$   0.036&  10.5 $\pm$   3.0 \\
 Gaia\_6151 & 1619258337664073088&  -2.475&   1.972 $\pm$   0.033&   0.798 $\pm$   0.042&   9.7 $\pm$   2.6 \\
 Gaia\_54 & 1235633782930379776&  -2.181&   0.819 $\pm$   0.014&   0.765 $\pm$   0.048&   6.4 $\pm$   4.6 \\
 Gaia\_1120 & 1763032952014500096&  -1.905&   0.356 $\pm$   0.006&   0.674 $\pm$   0.025&   5.6 $\pm$   4.4 \\
 Gaia\_42 & 1992247007192756224&  -1.572&  31.110 $\pm$   0.525&   0.831 $\pm$   0.031&  11.2 $\pm$   1.3 \\
 Gaia\_42 & 1992247007192756224&  -1.595&  31.110 $\pm$   0.525&   0.831 $\pm$   0.031&  11.2 $\pm$   1.3 \\
 Gaia\_66 & 2033964849204648064&  -0.648&   0.977 $\pm$   0.016&   0.834 $\pm$   0.057&   9.8 $\pm$   3.5 \\
 Gaia\_53 & 2038062247987954048&  -1.758&   3.054 $\pm$   0.052&   0.788 $\pm$   0.030&  12.5 $\pm$   1.6 \\
 Gaia\_11 & 1953616147184247808&  -1.558& 113.432 $\pm$   1.914&   0.834 $\pm$   0.034&  11.1 $\pm$   1.4 \\
 Gaia\_646 & 4510483566596988416&  -0.378&   6.023 $\pm$   0.102&   1.268 $\pm$   0.165&   3.8 $\pm$   1.0 \\
 Gaia\_32 & 1413748207000103808&  -2.004&   0.399 $\pm$   0.007&   0.688 $\pm$   0.028&   5.5 $\pm$   4.4 \\
 Gaia\_55 & 1420093656106274176&  -0.611&   1.216 $\pm$   0.021&   0.863 $\pm$   0.056&   9.8 $\pm$   3.6 \\
 Gaia\_1 & 1353777750444932736&  -1.294&  71.310 $\pm$   1.204&   0.844 $\pm$   0.033&  11.5 $\pm$   1.5 \\
 Gaia\_47 & 2145301671471881984&  -1.330&  46.579 $\pm$   0.786&   0.935 $\pm$   0.027&   7.8 $\pm$   0.6 \\
 Gaia\_1220 & 4250930137323602944&  -1.790& 163.637 $\pm$   2.762&   0.838 $\pm$   0.040&  10.7 $\pm$   1.8 \\
 Gaia\_2799 & 946807746355535360&  -0.771&   3.396 $\pm$   0.057&   1.000 $\pm$   0.094&   7.3 $\pm$   1.6 \\
 Gaia\_2082 & 187566551874865664&  -0.181&   4.006 $\pm$   0.068&   1.093 $\pm$   0.112&   5.4 $\pm$   1.4 \\
 Gaia\_26 & 1814359288672674560&  -0.962& 409.346 $\pm$   6.909&   0.929 $\pm$   0.027&  10.8 $\pm$   0.8 \\
 Gaia\_2620 & 3069023032306067968&  -1.597&  55.183 $\pm$   0.931&   1.069 $\pm$   0.035&   4.7 $\pm$   0.5 \\
 Gaia\_8 & 1768154507240342528&  -1.750&  60.189 $\pm$   1.016&   0.830 $\pm$   0.034&  11.1 $\pm$   1.4 \\
 Gaia\_3444 & 340605204521472640&  -0.944&   2.144 $\pm$   0.036&   0.880 $\pm$   0.076&  10.6 $\pm$   2.2 \\
 Gaia\_1288 & 2095628244392589824&  -1.708& 284.018 $\pm$   4.793&   0.826 $\pm$   0.026&  11.5 $\pm$   1.1 \\
 Gaia\_651 & 1907386149999121920&   0.015&   0.639 $\pm$   0.011&   0.939 $\pm$   0.039&   4.5 $\pm$   4.0 \\
 Gaia\_4892 & 1977970978277811840&  -1.041& 398.908 $\pm$   6.732&   0.927 $\pm$   0.024&  12.8 $\pm$   0.9 \\
 Gaia\_234 & 2228292706060074112&  -1.645& 382.410 $\pm$   6.454&   0.813 $\pm$   0.014&  13.3 $\pm$   0.7 \\
 Gaia\_57 & 4288765848587649664&  -0.589&   1.407 $\pm$   0.024&   0.933 $\pm$   0.075&   6.5 $\pm$   4.3 \\
 Gaia\_360 & 343680744701496448&  -1.068&   1.475 $\pm$   0.025&   0.837 $\pm$   0.067&  10.3 $\pm$   3.8 \\
 Gaia\_50 & 4500737499823974656&  -1.652&  52.292 $\pm$   0.883&   0.856 $\pm$   0.035&  10.4 $\pm$   1.4 \\
 Gaia\_5857 & 3779863003378779392&  -1.060&   2.331 $\pm$   0.039&   0.879 $\pm$   0.080&   9.4 $\pm$   2.8 \\
 Gaia\_3 & 1335426489060934272&  -1.888& 1055.064 $\pm$  17.806&   0.800 $\pm$   0.018&  12.2 $\pm$   0.9 \\
 Gaia\_16 & 1778657836877668096&  -1.352&  37.086 $\pm$   0.626&   1.605 $\pm$   0.054&   1.2 $\pm$   0.1 \\
 Gaia\_14 & 1984025890028941184&  -1.634&  43.450 $\pm$   0.733&   0.819 $\pm$   0.022&  11.8 $\pm$   1.0 \\
 Gaia\_2150 & 3644341873063382272&  -0.768&   1.407 $\pm$   0.024&   0.855 $\pm$   0.069&   8.9 $\pm$   3.2 \\
 Gaia\_46 & 1839321466796403072&  -1.142& 230.435 $\pm$   3.889&   0.827 $\pm$   0.023&  12.9 $\pm$   1.3 \\
 Gaia\_1671 & 1863896960443202816&  -1.656&   1.440 $\pm$   0.024&   0.791 $\pm$   0.048&   9.5 $\pm$   3.4 \\
 Gaia\_379 & 4606414581726715264&  -2.502& 206.256 $\pm$   3.481&   0.822 $\pm$   0.033&  11.1 $\pm$   1.5 \\
 Gaia\_23 & 1345432216672749184&  -1.575& 121.314 $\pm$   2.047&   1.246 $\pm$   0.038&   2.8 $\pm$   0.3 \\
 Gaia\_4550 & 2699001399207617792&  -1.079&  29.349 $\pm$   0.495&   0.837 $\pm$   0.030&  12.1 $\pm$   1.3 \\
 Gaia\_51 & 1776283372797846272&  -1.730& 205.049 $\pm$   3.461&   0.818 $\pm$   0.021&  11.7 $\pm$   0.9 \\
 Gaia\_28 & 4608346805911470080&  -2.861&  49.595 $\pm$   0.837&   1.530 $\pm$   0.053&   1.4 $\pm$   0.2 \\
 Gaia\_3033 & 3910720318902016000&  -2.602&   4.912 $\pm$   0.083&   0.785 $\pm$   0.018&  12.4 $\pm$   1.0 \\
 Gaia\_64 & 4205559241036808320&  -0.633&   0.564 $\pm$   0.010&   0.794 $\pm$   0.049&   7.8 $\pm$   4.5 \\
 Gaia\_4960 & 2668731259979103872&  -2.455&   2.625 $\pm$   0.044&   0.798 $\pm$   0.041&  10.6 $\pm$   2.3 \\
 Gaia\_1319 & 2047058074180427904&  -2.261&   3.093 $\pm$   0.052&   0.778 $\pm$   0.021&  12.4 $\pm$   1.3 \\
 Gaia\_1024 & 1421352528201172608&  -1.294&  92.496 $\pm$   1.561&   0.840 $\pm$   0.032&  11.5 $\pm$   1.3 \\
 Gaia\_2616 & 3070924882479137408&  -1.613&  31.118 $\pm$   0.525&   0.844 $\pm$   0.033&  10.9 $\pm$   1.4 \\
 Gaia\_36 & 4236626594855062656&  -1.586& 261.964 $\pm$   4.421&   0.834 $\pm$   0.041&  11.4 $\pm$   1.7 \\

 \hline\hline
    \end{tabular}
\end{table*}

\section{Mass and age of systems with low mass companions}

\begin{table*}[h]
    \caption{New derived individual masses and system age of stellar systems with low mass companions using \citet{maxted2025}.}
    \label{tab:lowmass_agemass}
    \centering
    \begin{tabular}{llcccc}
    \hline \hline
    Star & Gaia DR3 source ID & $q$ & $M_{\rm comp}$ & $M_{\rm primary}$ & 
    Age \\
    & & & [M$_\odot$] &  [M$_\odot$] & [Ga]\\
    
    \hline

EBLM\_J0057-19 & 2356249333811520512 &   0.126 &   0.133&   1.059 $\pm$   0.027&   8.3 $\pm$   1.1 \\
 EBLM\_J0113+31 & 313579414867698176 &   0.190 &   0.196&   1.034 $\pm$   0.018&   6.8 $\pm$   0.3 \\
 EBLM\_J0123+38 & 323261744326472832 &   0.262 &   0.342&   1.305 $\pm$   0.007&   5.0 $\pm$   0.1 \\
 EBLM\_J0228+05 & 2517108778307431168 &   0.119 &   0.171&   1.435 $\pm$   0.028&   1.3 $\pm$   0.2 \\
 EBLM\_J0500-46 & 4810245453195049472 &   0.162 &   0.169&   1.049 $\pm$   0.027&   7.1 $\pm$   0.7 \\
 EBLM\_J0526-34 & 4822911449189918208 &   0.252 &   0.323&   1.284 $\pm$   0.018&   3.8 $\pm$   0.1 \\
 EBLM\_J0540-17 & 2970781118606872192 &   0.140 &   0.176&   1.258 $\pm$   0.024&   4.1 $\pm$   0.3 \\
 EBLM\_J0608-59 & 5494443978353833088 &   0.282 &   0.324&   1.148 $\pm$   0.018&   4.7 $\pm$   0.3 \\
 EBLM\_J0627-67 & 5280290694161913472 &   0.246 &   0.282&   1.148 $\pm$   0.041&   3.6 $\pm$   0.4 \\
 EBLM\_J0719+25 & 870991952855398656 &   0.144 &   0.161&   1.118 $\pm$   0.018&   5.8 $\pm$   0.4 \\
 EBLM\_J0941-31 & 5440085875822885888 &   0.172 &   0.235&   1.363 $\pm$   0.023&   3.2 $\pm$   0.1 \\
 EBLM\_J0955-39 & 5419963580383767936 &   0.192 &   0.205&   1.071 $\pm$   0.026&   1.7 $\pm$   0.5 \\
 EBLM\_J1013+01 & 3835492798481111168 &   0.180 &   0.164&   0.915 $\pm$   0.019&  11.2 $\pm$   0.9 \\
 EBLM\_J1305-31 & 6180966394355426816 &   0.254 &   0.281&   1.107 $\pm$   0.013&   7.9 $\pm$   0.3 \\
 EBLM\_J1522+42 & 1393971390134144640 &   0.165 &   0.162&   0.983 $\pm$   0.015&  10.6 $\pm$   0.6 \\
 EBLM\_J1928-38 & 6739146911148825344 &   0.266 &   0.277&   1.041 $\pm$   0.015&   8.9 $\pm$   0.4 \\
 EBLM\_J1934-42 & 6689082332805585280 &   0.177 &   0.182&   1.024 $\pm$   0.017&   4.6 $\pm$   1.2 \\
 EBLM\_J2040-41 & 6678103674840655360 &   0.152 &   0.147&   0.969 $\pm$   0.015&  10.0 $\pm$   0.6 \\
 EBLM\_J2046+06 & 1735977265094166912 &   0.200 &   0.278&   1.394 $\pm$   0.021&   0.7 $\pm$   0.2 \\
 EBLM\_J2046-40 & 6678572680972611328 &   0.152 &   0.154&   1.008 $\pm$   0.008&  13.6 $\pm$   0.2 \\
 EBLM\_J2217-04 & 2626910437568266240 &   0.199 &   0.220&   1.105 $\pm$   0.013&   8.9 $\pm$   0.3 \\
 EBLM\_J2315+23 & 2839193907753424384 &   0.209 &   0.232&   1.113 $\pm$   0.015&   6.8 $\pm$   0.3 \\
 EBLM\_J2343+29 & 2867806567563054720 &   0.114 &   0.096&   0.842 $\pm$   0.015&   9.0 $\pm$   1.8 \\
 EBLM\_J2359+44 & 1923017735011519488 &   0.218 &   0.316&   1.453 $\pm$   0.018&   2.1 $\pm$   0.1 \\
 EPIC\_219654213 & 4183945728602365568 &   0.172 &   0.227&   1.319 $\pm$   0.031&   2.8 $\pm$   0.1 \\
 NGTS-EB-7 & 5504617415848984320 &   0.088 &   0.094&   1.067 $\pm$   0.022&   9.8 $\pm$   0.6 \\
 WASP-30 & 2434845586060285824 &   0.049 &   0.061&   1.249 $\pm$   0.026&   2.9 $\pm$   0.5 \\
 CoRoT-1 & 3105507886130792448 &   0.001 &   0.001&   1.206 $\pm$   0.012&   1.7 $\pm$   0.2 \\
 HAT-P-7 & 2129256395211984000 &   0.001 &   0.002&   1.543 $\pm$   0.009&   2.1 $\pm$   0.1 \\
 HD\_189733 & 1827242816201846144 &   0.001 &   0.001&   0.825 $\pm$   0.009&   4.4 $\pm$   0.9 \\
 HD\_209458 & 1779546757669063552 &   0.001 &   0.001&   1.145 $\pm$   0.009&   2.6 $\pm$   0.3 \\
 Kepler-1 & 2131314401800665344 &   0.001 &   0.001&   1.072 $\pm$   0.008&   1.4 $\pm$   0.3 \\
 WASP-4 & 6535499658122055552 &   0.001 &   0.001&   0.941 $\pm$   0.010&   5.9 $\pm$   0.8 \\
 WASP-18 & 4955371367334610048 &   0.007 &   0.010&   1.326 $\pm$   0.009&   0.2 $\pm$   0.1 \\
 WASP-43 & 3767805209112436736 &   0.003 &   0.002&   0.658 $\pm$   0.014&  11.5 $\pm$   2.3 \\
 KOI-205 & 2077896798333728640 &   0.042 &   0.039&   0.921 $\pm$   0.028&   3.4 $\pm$   2.1 \\
 
 \hline\hline
    \end{tabular}
\end{table*}

\section{Mass and age of scvPIC stars}

\begin{table}[h]
    \caption{New derived masses and ages of a subset of the scvPIC \citep{zwintz26}}     \label{tab:pic}
    \centering
    \begin{tabular}{lcccc}
    \hline \hline
    Gaia DR3 source ID  & $M_{\odot}$ & Age \\
     &   [M$_\odot$] & [Ga]\\
\hline
 5318186221414047104&   0.977 $\pm$   0.013&   4.6 $\pm$   0.9 \\
 5317884439832479872&   0.927 $\pm$   0.048&   7.5 $\pm$   3.8 \\
 5299155152615066752&   1.234 $\pm$   0.008&   2.7 $\pm$   0.2 \\
 5537152725129555712&   0.804 $\pm$   0.010&  13.1 $\pm$   0.7 \\
 5537149289155845888&   0.959 $\pm$   0.005&   8.5 $\pm$   0.3 \\
 4823123689293923968&   0.903 $\pm$   0.027&  10.8 $\pm$   2.3 \\
 4777119126354782592&   0.766 $\pm$   0.004&  13.4 $\pm$   0.4 \\
 4791496409119237888&   0.741 $\pm$   0.013&  12.4 $\pm$   1.2 \\
 5538812506654586624&   1.230 $\pm$   0.037&   1.6 $\pm$   0.7 \\
 5558725074543077632&   1.260 $\pm$   0.041&   2.3 $\pm$   0.5 \\
    \hline
 \hline\hline
    \end{tabular}
    \tablefoot{The full table of 339 stars is available online.}
    
\end{table}

\end{document}